\documentclass[preprint,flushrt,a4]{aastex}

\usepackage[varg]{txfonts}
\usepackage{graphicx, natbib, amssymb}
\bibpunct{(}{)}{;}{a}{}{,}

\newcommand{\kms}{km~s$^{-1}$}

\shorttitle{ISN~He haze}
\shortauthors{Sok{\'o}{\l} et al. 2015}

\begin{document}

\title{The Interstellar Neutral He haze in the heliosphere: what can we learn?}
    
\author{J.~M.~Sok{\'o}{\l}\altaffilmark{1}, M.~Bzowski\altaffilmark{1}, M.~A.~Kubiak\altaffilmark{1}, P.~Swaczyna\altaffilmark{1}, A.~Galli\altaffilmark{2}, P.~Wurz\altaffilmark{2}, E.~M{\"o}bius \altaffilmark{3}, H.~Kucharek \altaffilmark{3}, S.~A.~Fuselier\altaffilmark{4,5}, D.~J.~McComas\altaffilmark{4,5}}

\altaffiltext{1}{Space Research Centre of the Polish Academy of Sciences, 00-716 Warsaw, Poland}
\altaffiltext{2}{Physics Institute, University of Bern, Bern 3012, Switzerland}
\altaffiltext{3}{University of New Hampshire, Durham, NH 03824, USA}
\altaffiltext{4}{Southwest Research Institute, San Antonio, TX 78228, USA}
\altaffiltext{5}{University of Texas, San Antonio, TX 78249, USA}

\begin{abstract}
Neutral interstellar helium has been observed by the \emph{Interstellar Boundary Explorer (IBEX)} since 2009 with a signal-to-noise ratio well above 1000. Because of the geometry of the observations, the signal observed from January to March each year is the easiest to identify. However, as we show via simulations, the portion of the signal in the range of intensities from $10^{-3}$ to $10^{-2}$ of the peak value, previously mostly left out from the analysis, may bring important information about the details of the distribution function of interstellar He gas in front of the heliosphere. In particular, these observations may inform us about possible departures of the parent interstellar He population from equilibrium. We compare the expected distribution of the signal for the canonical assumption of a single Maxwell--Boltzmann population with the distributions for a superposition of the Maxwell--Boltzmann primary population and the recently discovered Warm Breeze, and for a single primary population given by a kappa function. We identify the regions on the sky where the differences between those cases are expected to be the most visible against the background. We discuss the diagnostic potential of the fall peak of the interstellar signal, reduced by a factor of 50 due to the Compton--Getting effect but still above the detection limit of IBEX. We point out the strong energy dependence of the fall signal and suggest that searching for this signal in the data could bring an independent assessment of the low-energy measurement threshold of the \emph{IBEX}-Lo sensor.
\end{abstract}


\section{Introduction}
Considerable progress has recently been made in the experimental studies of interstellar neutral (ISN) gas, especially helium. This progress was possible due to observations with the \emph{IBEX}-Lo instrument \citep{fuselier_etal:09b} on the \emph{Interstellar Boundary Explorer} \citep[\emph{IBEX;}][]{mccomas_etal:09a}. After the successful detection of ISN~He by \citet{mobius_etal:09b}, the first round of analysis of ISN~He observations by \citet{bzowski_etal:12a} and \citet{mobius_etal:12a} suggested that the ISN~He inflow parameters may be different than previously thought \citep{mobius_etal:04a}. The \emph{IBEX}-Lo observations from the 2009 and 2010 ISN observation seasons were consistent either with the gas flowing slower by $\sim 3$~\kms and from a direction different by $\sim 4\degr$ than that inferred from \emph{GAS/Ulysses} observations \citep{witte:04}, but having the same temperature, or alternatively, having a similar flow vector but with the temperature being considerably higher \citep{mccomas_etal:15a, mobius_etal:15a}. While these dilemmas have not been fully resolved, reanalysis of the earlier \emph{Ulysses} observations (including interpretation of the previously not analyzed last observation season 2007) by \citet{bzowski_etal:14a}, as well as a recent analysis of the \emph{IBEX} observations from 2013 and 2014 \citep{mccomas_etal:15a}, suggest that the second interpretation may be more likely.

\emph{IBEX}-Lo observations revealed that in addition to the expected ISN~He flow, another population inflowing from beyond the heliopause is present, dubbed the Warm Breeze \citep[WB;][]{kubiak_etal:14a}. The WB is flowing from a direction different by $\sim 20\degr$ (mostly in ecliptic longitude) from the inflow direction of ISN~He, and is approximately half as fast, and almost three times warmer than the primary ISN~He. In the initial study, this flow was analyzed as another homogeneous Maxwell--Boltzmann population directly upstream of the heliopause. \citet{kubiak_etal:14a} discussed various physical scenarios possibly responsible for the WB. These scenarios must result in considerable departures of this population from a Maxwell--Boltzmann distribution, also in terms of the spatial homogeneity. The kappa distribution of ISN~He in front of the heliopause was considered as one of the alternative possibilities of the source mechanism for the Warm Breeze. 

Kappa distributions seem to form naturally in cosmic plasmas \citep[see the recent review by][]{livadiotis_mccomas:13a}. Because of the elastic and charge-exchange collisions of ions from the ambient plasma with the ambient neutral component, including He atoms, departures from the ideal equilibrium state should be transmitted into the neutral population, and while not necessarily forming another kappa distribution, they certainly should be different from an ideal Maxwell--Boltzmann. Estimates by \citet{kubiak_etal:14a} suggested that the hypothesis that the source population for the WB is a kappa distribution should be rejected, but the hypothesis that the original ISN~He population is kappa-like rather than Maxwell--Boltzmann-like could not be ruled out. The departures of the ISN~He flux in the heliosphere from Gaussianity were also studied by \citet{gruntman:86a, gruntman:13a}, who considered elastic collisions of ISN~He atoms with solar protons. Such collisions should produce a halo, approximately $30\degr$-wide, around the maximum of the core of the flux.

On the experimental side, \citet{fuselier_etal:14a} and \citet{galli_etal:14a} studied the sources of the foreground and background of the \emph{IBEX}-Lo heliospheric signal and noticed a weak signal of $0.0005 \pm 0.0002$ times the maximum ISN signal strength (the lower end of the curve in Figure~\ref{figDataContrast}) in the four lower energy bins beside the ISN inflow and the WB, more or less uniformly distributed on the sky. This ubiquitous background appears in all energies from 15 to 110~eV and is approximately $50\%$ of the heliospheric signal at 150~eV. It persists along the whole Earth's orbit and in all directions, and does not exhibit a Compton--Getting effect \citep{mccomas_etal:10c}, which indicates a local origin \citep{galli_etal:14a}. Thus, background arising from the terrestrial magnetosphere or some effect internal to the instrument would seem to be the most likely explanations.

In this paper, which is a part of a coordinated set of Special Issue papers on ISN atoms measured by \emph{IBEX}, for which an overview is provided by \citet{mccomas_etal:15b}, we suggest possible experimental tests to provide a better understanding of the nature of interstellar populations of neutral He inside the heliosphere. As we will demonstrate, it seems that the best visible differences between scenarios with different distribution functions of the gas in the source region should be manifested at the edges of the core of ISN~He beam, observed by \emph{IBEX} in the spring of each year at Earth's ecliptic longitude $\lambda_{E}\sim 136\degr$. These wings of the parent distribution in the interstellar medium should form a weak, distributed He atom flux on the sky, which we call ``haze'' due to its similarity to a very fine cloud of gas, covering a large portion of the space around the observer. We simulate it for the three lowest-energy channels for three different scenarios: first, a single-population Maxwell--Boltzmann flow; second, two Maxwell--Boltzmann flows: the ISN~He and the WB; and third, a single kappa population featuring extreme departures from equilibrium, manifested by the low $\kappa$ index of $8/5$. Based on these simulations, we suggest the best times and locations on the sky to look for the signatures specific to the different scenarios.

Finally, we note that the results of this study are also important for future measurements of ISN atoms from spacecraft beyond \emph{IBEX}. In particular, the National Research Council (NRC -- an arm of the United States National Academies) recently completed ``The 2013-2022 Decadal Survey for Solar and Space Physics (Heliophysics)'', which defined an Interstellar Mapping and Acceleration Probe (\emph{IMAP}) as the next Solar Terrestrial Probe mission for NASA's Heliophysics Division. The planned IMAP payload includes a low energy $\left(\sim 5-1000 \mathrm{eV}\right)$ neutral atom camera to measure the inflowing H, D, He, O, and Ne with much higher sensitivity and angular resolution than possible on \emph{IBEX}. Such high-precision measurements of He (and other species) will carry on from \emph{IBEX} and even more strongly constrain models of the ionization state and radiation environment of LISM as well as uncover secondary populations and their detailed distribution functions. 

\section{Experimental aspects of \emph{IBEX}-Lo observations}
The \emph{IBEX} mission \citep{mccomas_etal:09a} comprises a spin-stabilized satellite that orbits the Earth in a highly elliptical orbit \citep{mccomas_etal:11a}. The \emph{IBEX}-Lo detector \citep{fuselier_etal:09b} is a time-of-flight mass spectrometer with the aperture pointing perpendicular to the spin axis. During each orbit or orbital segment, the aperture scans a single $\sim 7\degr$ FWHM wide strip of the sky (for details of the peaked collimator point-spread function see \citet{sokol_etal:15b}, this volume). The spin axis of the spacecraft is maintained within a few degrees from the Sun and very close to the ecliptic plane. Therefore, after half a year of single strip scans, the entire sky is covered. Polar regions are observed almost continuously, but the individual ecliptic sectors are observed just twice per year. The instrument registers neutral atoms in eight logarithmically spaced energy steps with the energy resolution $\Delta E/E \simeq 0.7$. In this paper we focus on the three lowest-energy channels, with center energies 14.5 (E1), 28.5 (E2), and 55.5~eV (E3). We also investigated the so-called energy step zero (E0), where we integrate over the whole energy spectrum.

\emph{IBEX}-Lo is sensitive to neutral helium via an indirect detection mechanism in which neutral He atoms impact on a conversion surface that is especially designed for surface ionization \citep{wurz_etal:06a, wieser_etal:07a}. This surface is covered with a thin layer comprised mostly of water, established by the outgassing of the instrument \citep{wurz_etal:09a}. The impact of He atoms sputters H, O, and C atoms and ions from the conversion surface, which are subsequently registered by \emph{IBEX}-Lo \citep{wurz_etal:08a}. Since the impact of the He atom on the surface is at a shallow angle, direct ejection of an atom from the surface layer by the impacting atom occurs following a process called recoil sputtering \citep{taglauer:90a}. The energy distribution of the sputtered H atoms is wide, ranging from below the impactor energy down to zero \citep{eckstein_etal:87a}, and is a weak function of the energy of the impacting He atom, except for a relatively narrow energy band immediately below the impactor energy \citep[for discussion see][]{mobius_etal:12a}. Thus, the signal of recoil-sputtered H$^{-}$ ions is visible in all energy steps below the energy of the impacting atoms. Since binding energies of the atoms on the surface have to be overcome by the recoil sputtering process, there is a minimum energy of the impactor required for the sputtering to occur, the so-called threshold energy \citep{taglauer:90a}, being the lower limit for the instrument energy range for He detection. An estimate for this energy was obtained by \citet[][this volume]{galli_etal:15a} based on simulations of the impact process.  The existence of an energy threshold for the observations of \emph{IBEX}-Lo has been qualitatively observed by \citet{kubiak_etal:14a}. They found that the behavior of the simulated signal at the wings of the WB beam is sensitive to the energy threshold and that the adoption of a finite threshold energy in the simulations improves the fit of the model to the data. However, details of this effect are not well known because reproducing the actual operational environment of the instrument in the laboratory is practically impossible. In this paper we propose an approximate experimental method of determining this sensitivity limit. 
	\begin{figure}
	\includegraphics{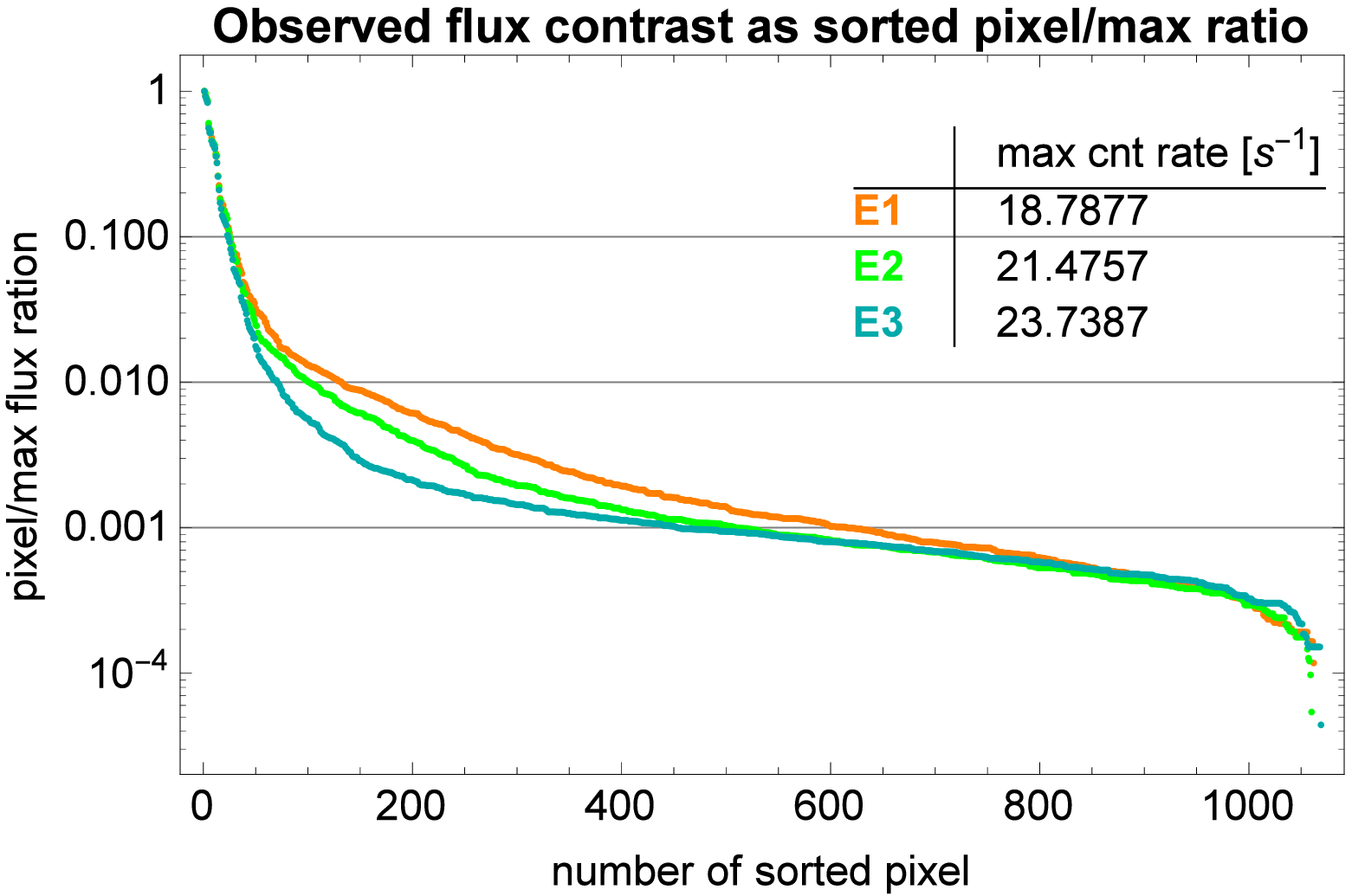}
	\caption{Ratios of count rates in individual \emph{IBEX}-Lo $6\degr$-bins (pixels) to the maximum count rate for a given energy step, sorted in increasing order, for all $6\degr$-pixels in energy steps 1 (orange), 2 (green), and 3 (blue) from orbits 54 through 72, shown as an example to illustrate the high dynamic range of \emph{IBEX}-Lo observations. The inset table shows the maximum count rates in all three energy steps, observed during the 2009/2010 ISN observation season. There are $18 \times 60 = 1080$ bins for one energy step. The bins from E2 constitute pixels of the data map shown further on in Figure~\ref{figMapE2obsWhite}. The background has not been subtracted.}
	\label{figDataContrast}
	\end{figure}

\emph{IBEX}-Lo data for the ISN~He flow have routinely achieved an extremely high contrast (the season maximum to minimum ratio) greater than $10^3$. The distribution of pixel brightness as shown in Figure~\ref{figDataContrast} has been repeated in all seasons thus far. The steep decrease at the left side of the plot is due to the ISN~He peak. The region between $\sim 10^{-3}$ and $\sim 10^{-2}$ is the main area of interest in this paper: it contains most of the WB and ISN distributed flux pixels, in addition to the ubiquitous background signal \citep{fuselier_etal:14a, galli_etal:14a}. The background signal seems to be in the regions below $\sim 7 \times 10^{-4}$ of the maximum value, where the signals from the all three energy steps start to diverge (i.e., starting from approximately the 800$^{th}$ pixel in the plot). For all three energy steps, a signal that is useful for the analysis of the darkest pixels is consistently as low as a few times $10^{-4}$ of the seasonal maximum. The effective observation times for a given pixel and energy step were not equal for the pixels shown in the figure; they varied from orbit to orbit from just a few minutes to a few hours per pixel per orbit. These times may seem short for an orbit of $\sim 7$ days, but one must keep in mind that the energy setting is stepped over the eight energy steps and that the data are collected in 60 $6\degr$-angular bins (pixels), which results in a $60 \times 8 = 480$-fold reduction in the effective observation time per bin compared with the total observation time for an orbit.

\begin{figure}
	\resizebox{\hsize}{!}{\includegraphics{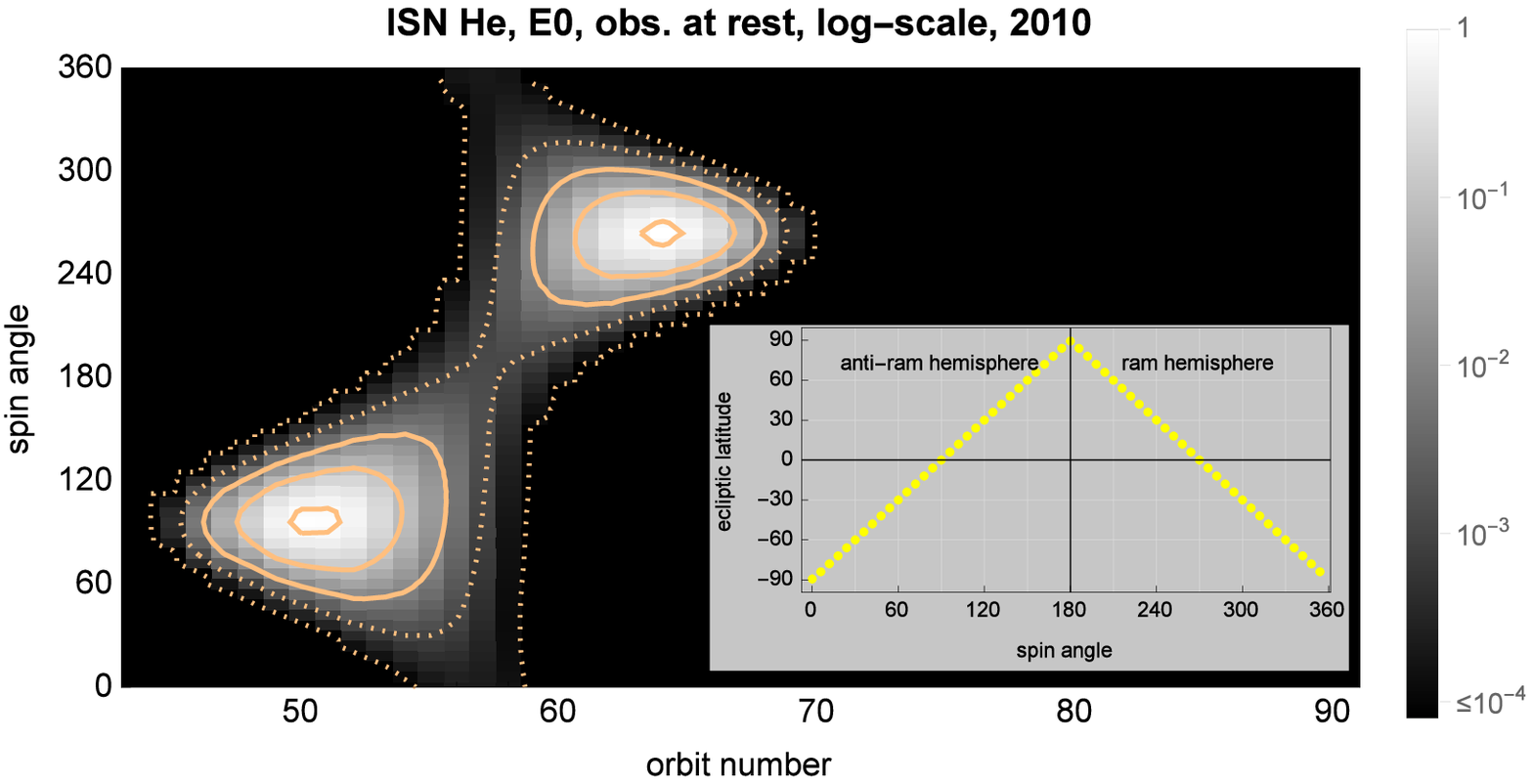}}
	\caption{Simulated full-sky map of the ISN~He flux (single Maxwell--Boltzmann population) as it would be seen by a virtual \emph{IBEX}-Lo located at ecliptic longitudes corresponding to the longitudes of \emph{IBEX} orbits 43 through 91 (2009-2010 to August 30), with spin axis oriented identically as it was in reality, but at rest in the Sun frame. The flux is shown in the logarithmic scale and rescaled so that the peak value is equal to 1. The solid isocontour lines show 0.8, $10^{-1}$, and $10^{-2}$ of the peak value, and the two broken isocontours correspond to $10^{-3}$ and $10^{-4}$ of the maximum flux, i.e., the isocontours cover (with some excess) the whole dynamical range of \emph{IBEX}-Lo data. The map is shown in the spacecraft reference system. The figure is built with vertical strips composed of $6\degr$-bins in spin-angle. Those strips correspond to the flux calculated for individual orbits for the full range of spin-angles, i.e., from $0\degr$ to $360\degr$, tabulated for the center of the actual $6\degr$-bins of \emph{IBEX}-Lo. The approximate correspondence between spin-angles and ecliptic latitude, in the convention adopted in this paper, is shown in the inset. A similar format is used in other sky maps presented in the paper.}
	\label{figMapE00nisVsc0}
	\end{figure}

If the parent distribution of the gas in front of the heliopause is homogeneous, the flow of the gas inside the heliosphere will feature axial symmetry, unless there are asymmetric ionization losses in the heliosheath or inside the heliosphere due to a heliolatitude dependence of the solar factors (for a discussion of the latter, see \citet{bzowski_etal:13a}). Therefore, the ISN~He signal observed by a non-moving observer along the Earth's orbit should be almost symmetric relative to the projection of the inflow direction on the ecliptic plane, as illustrated in Figure~\ref{figMapE00nisVsc0}. However, the \emph{IBEX} spacecraft is moving with the Earth, which causes an important difference as presented in Figure~\ref{figMapE00nis}. In spring ($\lambda_{E}$ from $\sim 60\degr$ to $\sim 170\degr$) the velocities of the spacecraft and the flow add, thus strongly enhancing the apparent flux via the Compton--Getting effect (modification of the magnitude of the registered flux by the relative motion of the flowing gas and the detector) and additionally increasing the measured count rate as the instrument sensitivity increases with increasing energy of neutral atoms entering the detector. In the fall ($\lambda_{E}$ complementary to the spring season), the situation is the opposite: the flux in the peak is reduced by almost two orders of magnitude due to subtracting the velocities of the ISN flow and the spacecraft. The effect of modification of the observed flux due to the relative velocity of the gas atoms and detector can be appreciated by comparing Figure~\ref{figMapE00nisVsc0} with Figure~\ref{figMapE00nis}. It is important to realize that even if the instrument had been equally sensitive at all energies, the fall peak must be quite reduced solely due to the difference in relative velocities of the gas and the instrument. Indeed, the modification of the trajectories of the atoms by solar gravity is symmetric with respect to the Sun, and for an assumed symmetry of the ionization rate, the ionization losses of the ISN~He gas on both sides of the Sun at a given time are identical. Thus, when the spring peak exists, the fall peak must also exist. The question is if it can be observed given the instrumental and observational constraints.

Once the parameters of the flow in front of the heliopause are known, it is straightforward to calculate the expected location in Earth's orbit where the spacecraft will observe the fall peak, and the apparent direction of the flux maximum.  While the absolute magnitudes of the spring and fall peaks depend on the ionization rate, the ratio of the peak maxima does not because the ionization losses on both sides of the Sun are nearly identical.\footnote{In reality, a small difference may exist because of the changes in the ionization rate with time.} With the background level established and with the known peak intensity for the spring peak, one can infer the expected intensity and location of the fall peak. If the peak is not observed, then the only reason (excluding a high local foreground) may be instrumental. In the following sections of the paper, this fact can be used to establish the practical sensitivity threshold of \emph{IBEX}-Lo for low-energy He atoms. 

\section{Adopted distribution functions and integration boundaries}
We carried out numerical simulations of the ISN~He flux as it would be observed by \emph{IBEX}-Lo on orbits 43 (2009 August 30) through 91 (2010 August 30), providing a full coverage of the Earth's orbit around Sun. The simulations aimed at reproducing several alternatives for the physical state of ISN~He gas in front of the heliopause, including an isotropic single Maxwell--Boltzmann, an isotropic single kappa population with a low $\kappa$ value of $8/5$ and several alternative values of the reference speed, and a superposition of two isotropic Maxwell--Boltzmann populations: the primary ISN~He gas and the WB, all normalized by the density of the ISN gas in the source region. 

For the Maxwell--Boltzmann distribution function, we adopted the conventional definition:
	\begin{equation}
	f_\mathrm{M-B}\left( v, u_\mathrm{M-B} \right) = \pi^{-3/2} u_\mathrm{M-B}^{-3} \exp \left[ -\frac{v^2}{u_\mathrm{M-B}^2} \right]; 
	\label{eqDefMaxwellBoltzmann}
	\end{equation}
for the kappa function we followed the definition from \citet{livadiotis_mccomas:13a} for which the fundamentals are extensively described in \citet{livadiotis_mccomas:09a, livadiotis_mccomas:11a}:
	\begin{equation}
	f_{\kappa} \left( v_, u_{\kappa} \right) =  \pi^{-3/2} u_{\kappa}^{-3} \left( \kappa-\frac{3}{2} \right)^{-3/2} \frac{\Gamma \left( \kappa + 1 \right)}{\Gamma \left( \kappa - \frac{1}{2} \right)}\left( \frac{v^2}{\left( \kappa -\frac{3}{2} \right) u_{\kappa}^2} +1 \right)^{-\kappa - 1},
	\label{eqDefKappa}
	\end{equation}
where $v$ is the atom speed in the reference frame comoving with the gas at the source region, the reference speed in the Maxwell--Boltzmann distribution $u_{\mathrm{lim}}$ is the thermal speed, and $u_{\kappa}$ is the reference speed in the kappa distribution function. 

Integration boundaries are an important aspect of our simulation. We integrate the flux in the spacecraft reference frame, but the atoms are transferred to the heliocentric frame and traced to the source region, located in front of the heliosphere. The integration goes from 0 relative speed (or from the speed corresponding to the adopted energy threshold) to infinity. In practice, however, as the upper boundary, we calculate a finite speed whose magnitude is determined from the conservation of energy and from the prerequisite that the calculation covers at least $99\%$ of the density of the neutral gas in the parent region. More information can be found in \citet[][this volume]{sokol_etal:15b}.

The latter condition is easily calculated for the Maxwell--Boltzmann distribution, where we must include all atoms that in the reference frame that comoving with the interstellar gas flow do not exceed $\sim2.4 u_{\mathrm{M-B}}$. With this limiting speed in the comoving frame, we turn to the solar inertial frame, in which the limiting speed is equal to $v_B + 2.4 u_{\mathrm{M-B}}$. Starting from this speed, we calculate the corresponding maximum speed at 1~AU from the conservation of energy. Then this speed is used to calculate the upper boundary in the spacecraft frame. Since this speed is relatively low, we increase this limit to $4.5 u_{\mathrm{M-B}}$, which leaves out only $\sim10^{-9}$ of the entire population. 

For the kappa function, we proceed similarly, i.e., we start out from the prerequisite that at least $1 - \Delta = 0.99$ of the population in front of the heliosphere is covered, and we calculate the upper speed boundary in the comoving frame. It can be shown that for a kappa distribution given by Equation~\ref{eqDefKappa} the absolute boundary speed $u_{\mathrm{lim}}$ is given by the following equation:
	\begin{equation}
	\Delta = 1 - \frac{4\Gamma \left( \kappa + 1 \right)}{3 \Gamma \left( \kappa - \frac{1}{2} \right)\left( \kappa - \frac{3}{2} \right)^{\frac{3}{2}}\pi^{\frac{1}{2}}} \left( \frac{u_{\mathrm{lim}}}{u_{\kappa}} \right)^3 {}_{2}F_{1}\left( \frac{3}{2}, \kappa + 1; \frac{5}{2}; \frac{1}{\frac{3}{2}-\kappa}\left( \frac{u_{\mathrm{lim}}}{u_{\kappa}} \right)^2 \right),
	\label{eqUlim}
	\end{equation}
where $\kappa$ is the parameter in the kappa distribution function defined in Equation~\ref{eqDefKappa}, and ${}_{2}F_{1}\left(a,b;c;z\right)$ is defined as	
	\begin{equation}
	{}_{2}F_{1}\left( a,b; c; z \right) = \Gamma \left( c \right) / \left[ \Gamma\left( b \right) \Gamma \left( c-b \right) \right]\int\limits_{0}^{1}{t^{b-1}\left( 1-t \right)^{c-b-1} \left( 1-tz \right)^{-a} \mathrm{d}t}.
	\label{eqDefHypergeometric}
	\end{equation}
	
\begin{figure}
	\resizebox{\hsize}{!}{\includegraphics{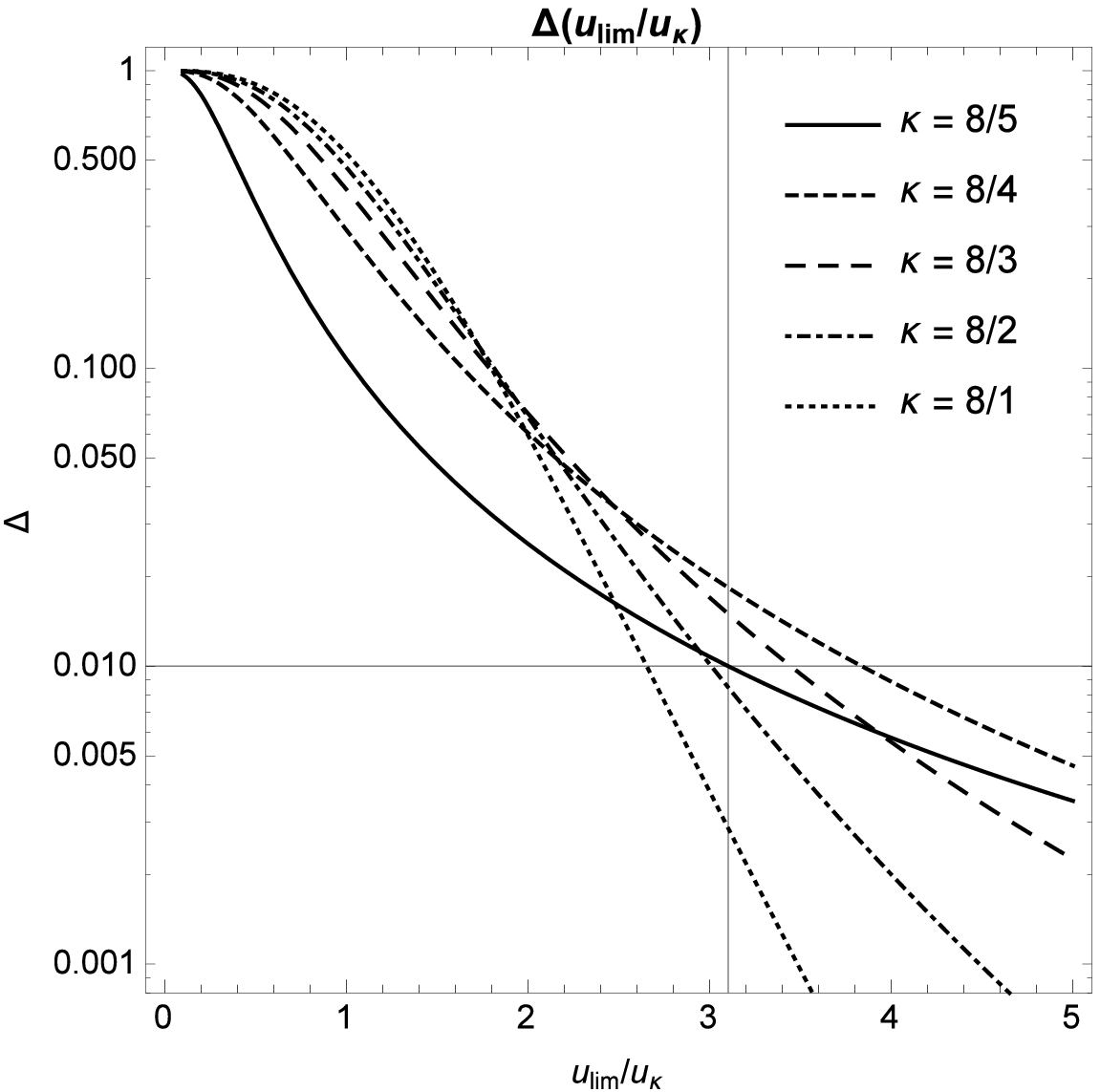}}
	\caption{Fraction $\Delta\left( u_{\mathrm{lim}}/u_{\kappa} \right)$ of the total density left out from integration of the kappa distribution function defined in Equation~\ref{eqDefKappa} from 0 to the upper boundary $u_{\mathrm{lim}}$. The crosshairs mark the $u_{\mathrm{lim}}/u_{\kappa}$ value for $\kappa = 8/5$ for the $\Delta\left( u_{\mathrm{lim}}/u_{\kappa} \right)=0.01$ adopted in the paper.}
	\label{figUlim}
	\end{figure}

Equation~\ref{eqUlim} is solved numerically for $u_{\mathrm{lim}}$ with the adopted value of $\kappa$. Solutions of this equation can be inferred from Figure~\ref{figUlim}, where the plots of $\Delta\left( u_{\mathrm{lim}}/u_{\kappa} \right)$ are provided for a few values of $\kappa$. For example, for $\kappa = 8/5$, $u_{\mathrm{lim}} = \sim 3.1 u_{\kappa}$, which for $u_{\kappa} = \sim20.4$~\kms gives $u_{\mathrm{lim}}\sim63$~\kms. Note that the boundary values $u_{\mathrm{lim}}/u_{\kappa}$ for various $\kappa$ are confined to a relatively narrow range from $\sim2.8$ to $\sim3.8$. 
	
\section{Numerical model and simulations}  
Simulations discussed in the paper were performed using a new and independent code called an analytic Warsaw Test Particle Model (aWTPM), described in detail by \citet{sokol_etal:15b}, developed based on assumptions and prerequisites similar to the Warsaw Test Particle Model used by, e.g.,~\citet[][]{tarnopolski_bzowski:08a, bzowski_etal:12a, bzowski_etal:14a}, and \citet{kubiak_etal:14a} in their studies. The program is implemented in the \texttt{Wolfram Research Mathematica} calculation system. It computes the expected ISN~He signal for an assumed distribution function in front of the heliopause with a given set of relevant parameters, taking into account the spacecraft velocity, the exact spin axis pointing, and the \emph{IBEX}-Lo collimator transmission function as described in \citet{sokol_etal:15b}. 

The aWTPM code used in this study follows the assumptions of the classical hot model \citep{thomas:78, fahr:79, wu_judge:79a}. The ionization rate is constant in time, but to alleviate the resulting inaccuracies, we adopted a quasi time-dependent approach, assuming, for each orbit, the instantaneous ionization rate resulting from the sum of photoionization, charge exchange, and electron impact ionization in the ecliptic plane for the time of detection with $1/r^2$ variations with the distance to Sun ($r$)\footnote{The ionization rates used in this calculations are released via IBEX data release in support of this ISN IBEX Special Issue of ApJS publication, Data Release~9 (see also \citet{sokol_etal:15b}).}. Among these the photoionization is the most effective; see \citet{bzowski_etal:13b}. The differences between the flux calculated using the stationary model with instantaneous ionization and a full time-dependent model are presented by \citet{rucinski_etal:03}, and they are not significant for the present study.

We calculated full-sky maps of the collimator-integrated ISN~He flux, tabulated every $6\degr$ in spin-angle for the whole range of spin-angles from $0\degr$ to $360\degr$. Note that this simplification is not fully equivalent to the flux integrated over $6\degr$ spin-angle bins, but is sufficient for the purpose of this study. Specifically, we calculated the flux for orbits 43 through 91, i.e., for the entire \emph{IBEX} orbit around the Sun, thus covering the entire sky in the ram (the detector points toward the direction of the velocity vector of the spacecraft) and anti-ram (the detector looks in the opposite direction than the velocity vector) hemispheres. The instrument sensitivities in different energy steps were not taken into account. For each orbit, we adopted the \emph{IBEX} spin-axis pointing \citep{swaczyna_etal:15a} to precisely determine the field of view. We also used the actual Earth velocity relative to the Sun and the spacecraft velocity relative to the Sun to precisely account for the relative speed of incoming He atoms. For ionization we used the photoionization model by \citet{sokol_bzowski:14a}, which is an update of the model by \citet{bzowski_etal:13b}, and which is based on the actual solar EUV spectral flux measured by TIMED \citep{woods_etal:05a}. The solar wind parameters for charge exchange and electron impact ionization from the ecliptic plane were taken from the OMNI database \citep{king_papitashvili:05}. For each orbit we calculated the ionization rate for the mid-time of the \emph{IBEX} High-Altitude Science Operations interval. The flux for a given orbit was calculated for this moment in time without integrating over the time of observations in the given orbit \citep{mobius_etal:12a, leonard_etal:15a}. The spin-angles from $0\degr$ to $180\degr$ cover the anti-ram hemisphere and the spin-angles from $180\degr$ to $360\degr$ cover the ram hemisphere (see the inset in Figure~~\ref{figMapE00nisVsc0}). 

\begin{figure}
	\resizebox{\hsize}{!}{\includegraphics{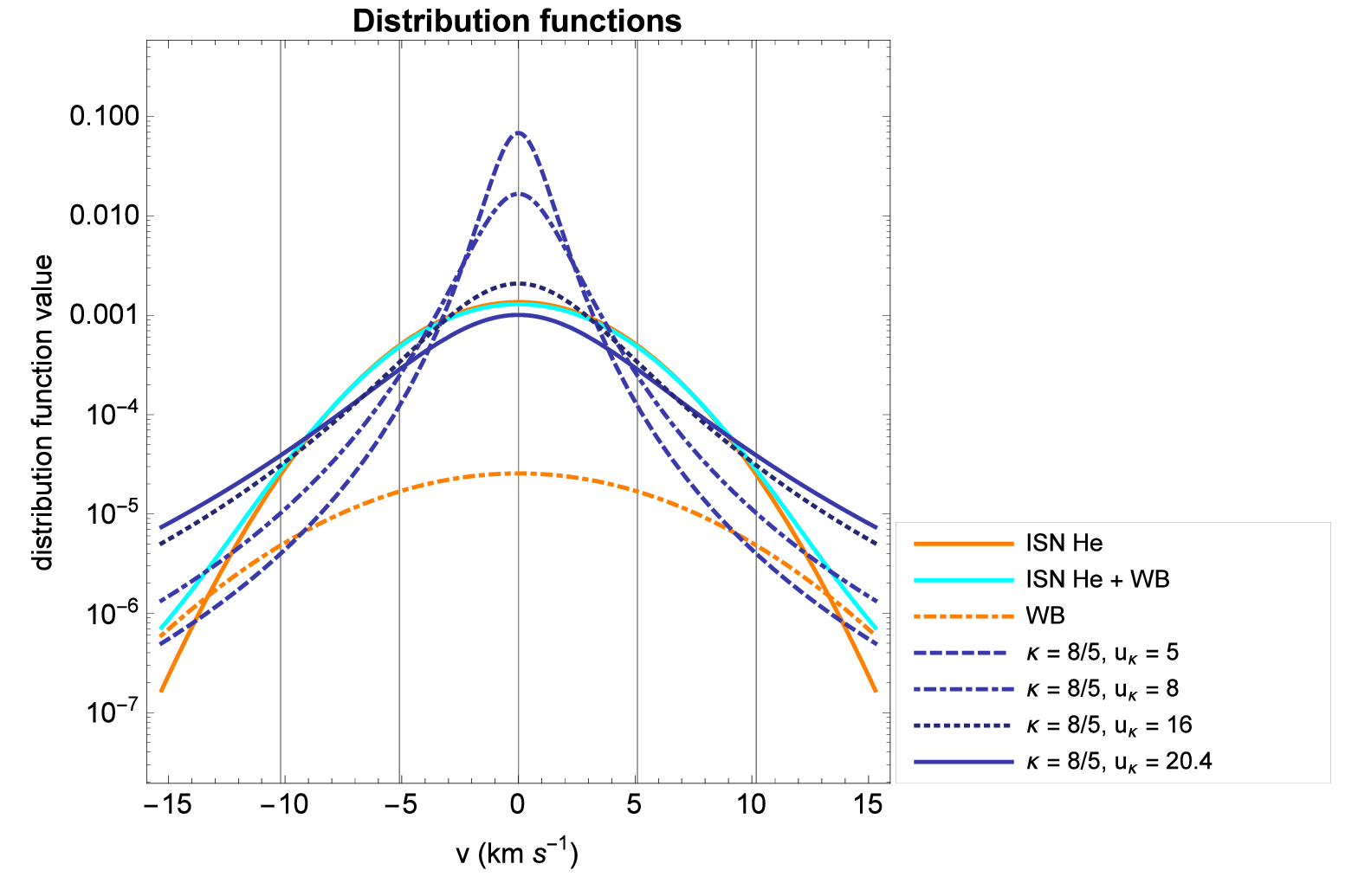}}
	\caption{Distribution functions used in this study: Maxwell--Boltzmann (Equation~\ref{eqDefMaxwellBoltzmann}) with thermal speed value $u_{\mathrm{lim}} =5.06$~\kms, characteristic for temperature 6165~K, as found by \citet{bzowski_etal:12a} for ISN~He population, Maxwell--Boltzmann with the thermal speed characteristic for the Warm Breeze temperature of 15068~K, as found by \citet{kubiak_etal:14a}. The density is scaled to maintain the proportions between the shown cases. The kappa distribution functions come from Equation~\ref{eqDefKappa} for $\kappa=8/5$ with different $u_{\kappa}$ given in~\kms. Note that the plot for the superposition of the ISN~He + Warm Breeze case is only an approximation because in reality the mean velocity vector of the Warm Breeze in the reference frame comoving with the ISN~He population is not zero. Thus, only the widths and absolute densities of the ISN~He and WB populations can be compared in this figure.}
	\label{figDistributionFunctions}
	\end{figure}

The simulations were performed for the ISN~He inflow parameters as reported by \citet{bzowski_etal:12a} and for the WB as given by \citet{kubiak_etal:14a}. The results are shown mostly for the 2009/2010 ISN observation season, but we checked that our conclusions do not critically depend on the choice of the year analyzed. We performed the simulations with the conventional assumption of a single Maxwell--Boltzmann population in front of the heliosphere, supplemented by the Maxwell--Boltzmann WB population, and alternatively, assuming that the ISN~He gas in front of the heliosphere features a kappa distribution, with the direction and bulk speed identical to that of the Maxwell--Boltzmann ISN~He population, but with $\kappa = 8/5 = 1.6$, and various reference speeds $u_{\kappa}$. We evaluated a few cases with $u_{\kappa}$ starting from $u_{\kappa}=5$~\kms to $u_{\kappa} = 20.4$~\kms. In the first case we get the same mean thermal energy as in the Maxwellian distribution we used, but the distribution function is much narrower, as shown in Figure~\ref{figDistributionFunctions}. The $u_{\kappa} = 20.4$~\kms was selected to obtain a similar shape of the distribution function close to the core of the distribution function in the rest frame, but the thermal energy in this distribution is much larger than in the Maxwell--Boltzmann distribution. However, since the shape of these two distributions are quite similar within $\sim \pm 5$\kms around the peak (see Figure~\ref{figDistributionFunctions}) we wanted to check the sensitivity of the observed flux to such differences in the parent population of the ISN~He gas. In addition, we simulated two intermediate cases, with $u_{\kappa}$ equal to 8 and 16~\kms. We chose the low kappa value ($\kappa = 8/5$) in order to test an extreme case and see if significant differences in the distribution of the haze flux are produced. An illustrative comparison of the distribution functions of all tested cases in the frame comoving with the interstellar gas is shown in Figure~\ref{figDistributionFunctions}. 

The results of the simulations are presented as full-sky maps of the logarithm of the flux, normalized to the flux maximum for the map (Figures~\ref{figMapE00nis}--\ref{figMapE00kappa5} and \ref{figMapE20niswb}--\ref{figMapE20kappa5}). In this way, the maximum of the flux on the map is always equal to 1, which facilitates direct comparison with observations. When one knows the actual count rate registered during a given season, it is possible to make assessments of the actual count rates expected in different regions of the map and for different cases. The normalized flux values are marked in grayscale, with isocontour lines drawn at $10^{-4}$, $10^{-3}$, $10^{-1}$, and 0.8 of the flux maximum.  

\section{Results}
\label{secResults}
\subsection{The full-sky ISN~He haze}
\begin{figure}
	\resizebox{\hsize}{!}{\includegraphics{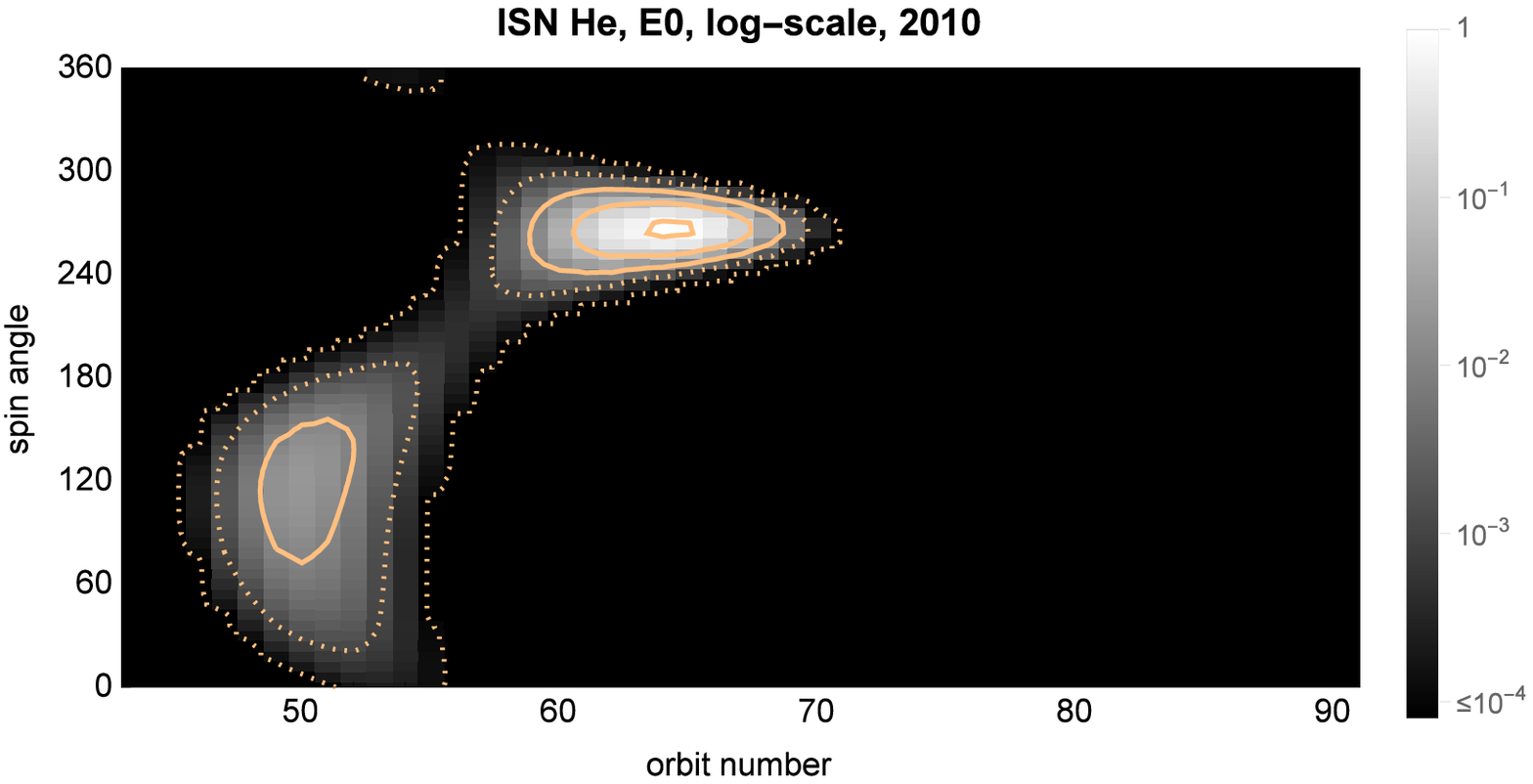}}
	\caption{Full-sky map of ISN~He flux for the single Maxwell-Boltzmann distribution (Equation~\ref{eqDefMaxwellBoltzmann}) of the primary ISN~He population as it would be seen by \emph{IBEX}-Lo, if its energy sensitivity was independent of energy. The thermal speed $u_{\mathrm{lim}}$ was $\sim5$~\kms. The map is constructed of vertical strips corresponding to the flux observed on subsequent orbits. The strips are located side by side, in chronological order. The flux is scaled to the global maximum (spring peak) and presented in the logarithm scale. The color ovals are the flux isocontours lines located at 0.8, $10^{-1}$,  $10^{-2}$ (solid lines), and  $10^{-3}$ and  $10^{-4}$ of the spring peak value (broken lines). The flux distribution on the sky for the values lower than $10^{-4}$ of the peak value are drawn as totally black.}
	\label{figMapE00nis}
	\end{figure}

\begin{figure}
	\resizebox{\hsize}{!}{\includegraphics{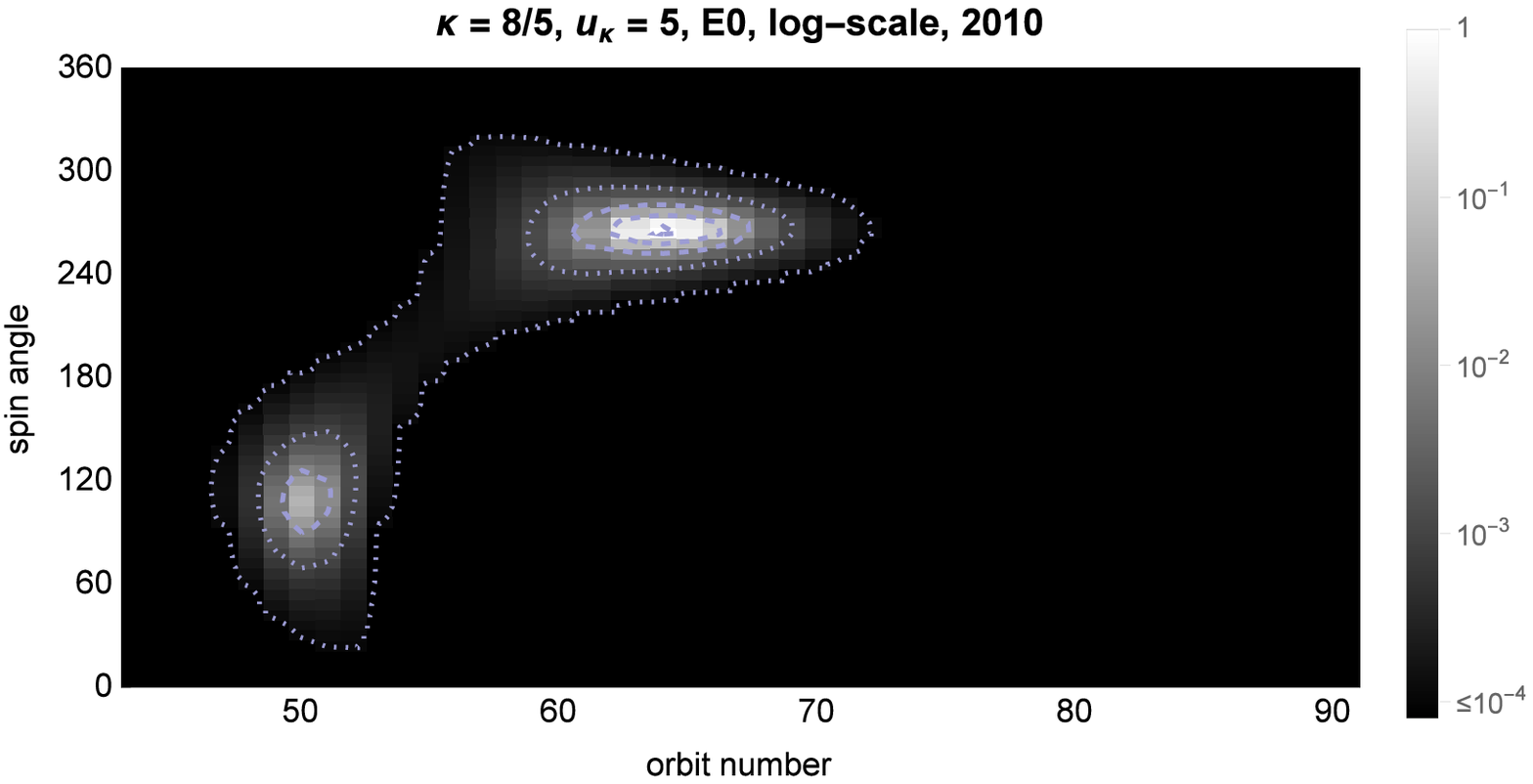}}
	\caption{Full-sky map of ISN~He flux for the narrow kappa distribution assumed in front of the heliopause, with $\kappa = 8/5$ and the reference speed $u_{\kappa} = 5$~\kms. The figure format is identical as in Figure~\ref{figMapE00nis}.}
	\label{figMapE00kappa1}
	\end{figure}

\begin{figure}
	\resizebox{\hsize}{!}{\includegraphics{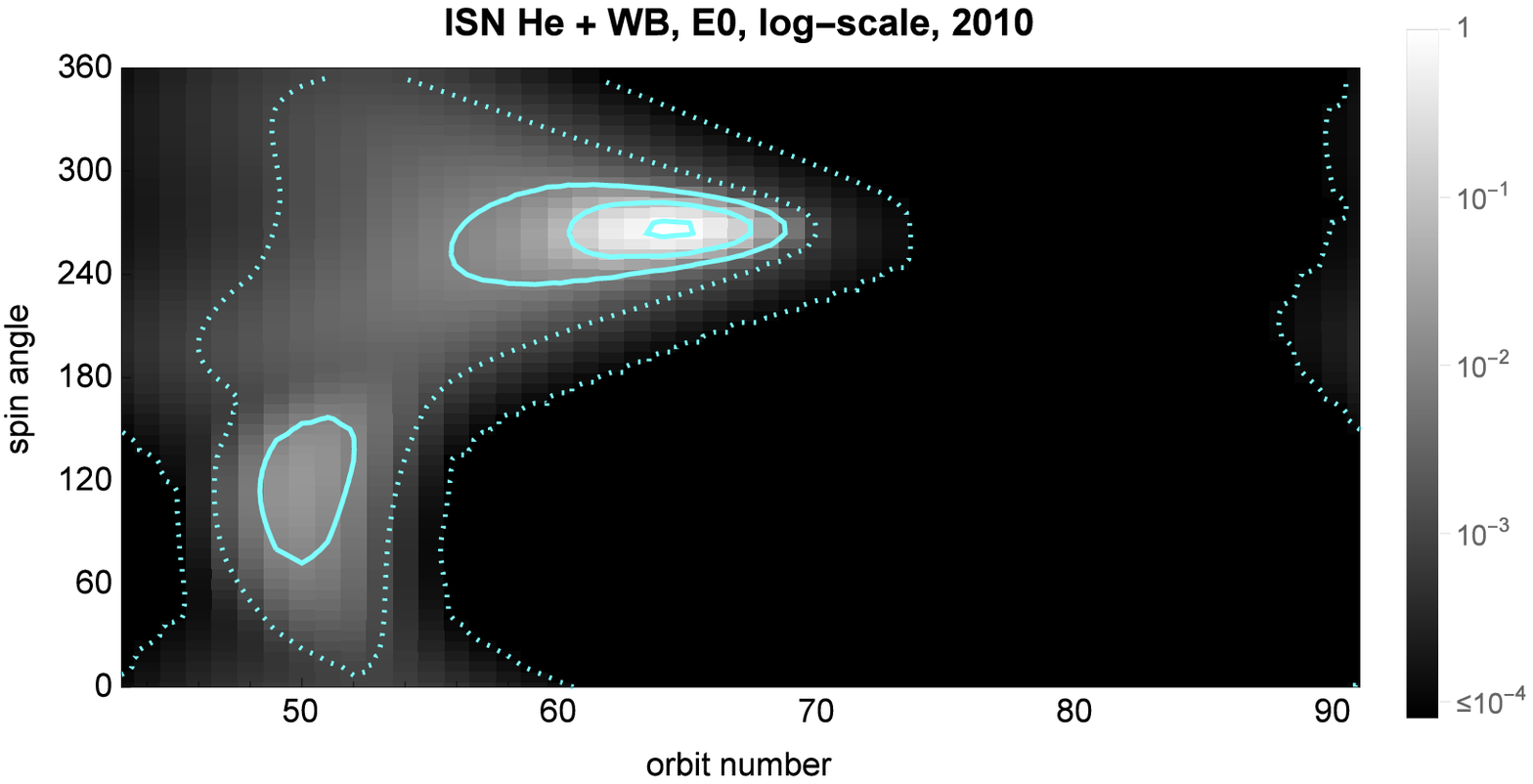}}
	\caption{Full-sky map of interstellar helium flux, simulated as a superposition of the flux of the ISN~He primary population, shown in Figure~\ref{figMapE00nis}, and of the Warm Breeze. The figure format is identical as to that in Figure~\ref{figMapE00nis}.}
	\label{figMapE00niswb}
	\end{figure}

\begin{figure}
	\resizebox{\hsize}{!}{\includegraphics{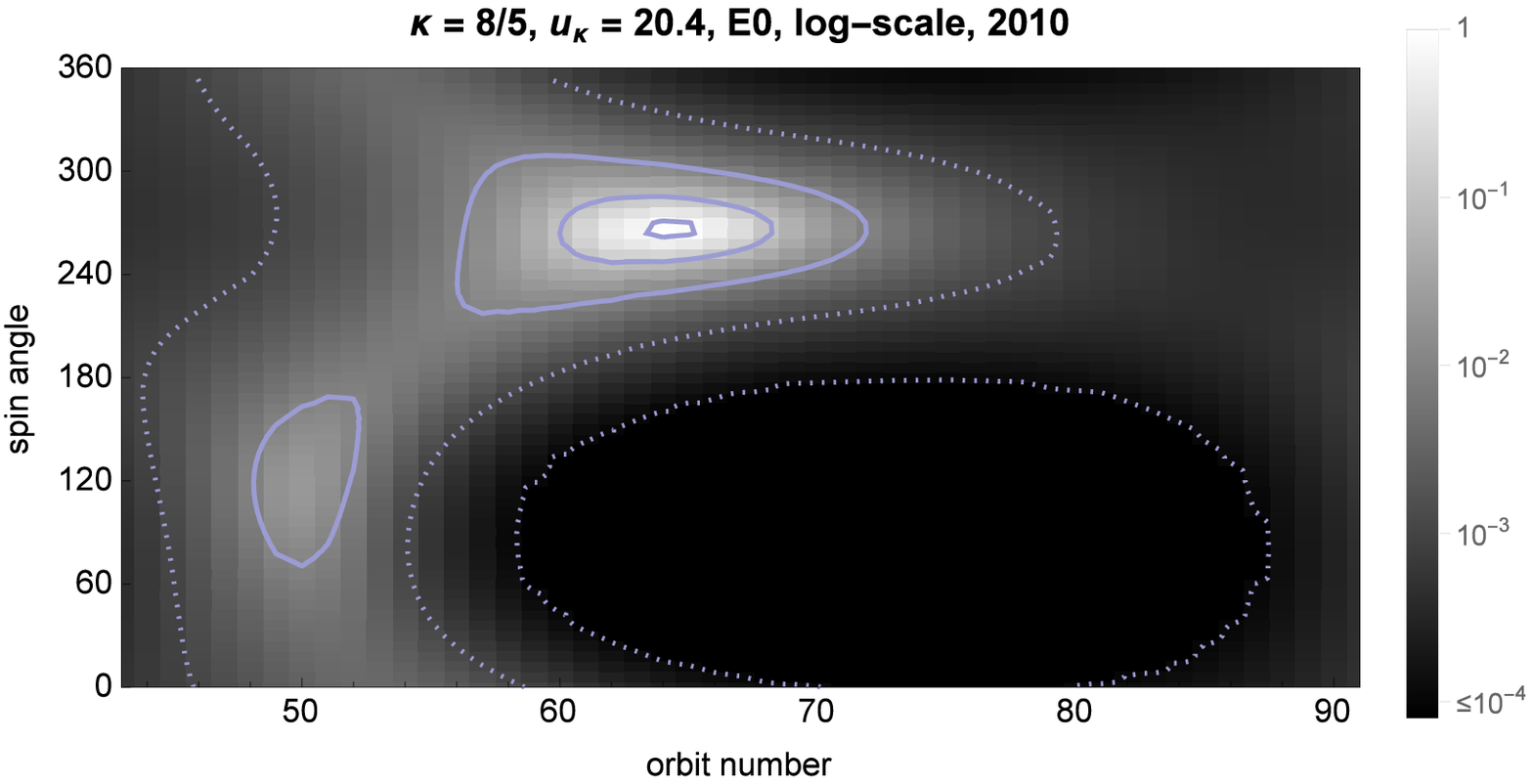}}
	\caption{Full-sky map of the ISN~He population simulated, assuming that the distribution function of ISN~He gas in front of the heliopause is given by the kappa function, with $\kappa = 8/5$ and the extreme case of the reference speed $u_{\kappa} = 20.4$~\kms. The figure format is identical as to that in Figure~\ref{figMapE00nis}.}
	\label{figMapE00kappa5}
	\end{figure}

Simulations performed for the full energy range, i.e., with no energy threshold (labeled as E0 in Figures~\ref{figMapE00nis}, \ref{figMapE00kappa1}, \ref{figMapE00niswb}, \ref{figMapE00kappa5}), show that regardless of the distribution function adopted in front of the heliopause: single or two-Maxwellian, or kappa function with various reference speed values, there are always two peaks of ISN~He expected, one in the ram hemisphere (the spring peak), and the other one in the anti-ram hemisphere (the fall peak). The fall peak should be visible in orbit 50 ($\lambda_{E}\sim 30\degr$) and the adjacent ones, as well as in equivalent orbits in all other years. The ratio of the fall peak to the spring peak heights is expected at $\sim0.02$, weakly dependent on the assumed distribution function and observation year. Figures~\ref{figMapE00nis} through \ref{figMapE00kappa5} show the maps with both peaks clearly visible.

Further inspection of Figures~\ref{figMapE00nis} through \ref{figMapE00kappa5} reveals that the differences between flux distributions on the sky for the various ISN~He distribution functions assumed in front of the heliosphere are most visible in the regions away from the spring peak, where the expected signal is relatively low. We refer to this extended region of low signals as the ISN~He haze in anticipation that this signal may be related to the foreground seen by \emph{IBEX}-Lo at low energies \citep{galli_etal:14a}. For the ISN~He Maxwellian population, the haze is very faint and almost the entire sky is dark, i.e., the signal is below $10^{-4}$ of the peak value, and certainly cannot be observed by \emph{IBEX} even with its high signal-to-noise ratio (Figure~\ref{figMapE00nis}). The same is true for a narrow kappa distribution with the reference speed $u_{\kappa}=5$~\kms (Figure~\ref{figMapE00kappa1}). In contrast, a superposition of the flux from the primary ISN~He population and the WB (Figure~\ref{figMapE00niswb}) occupies a large portion of the sky, mostly in the intensity range below 0.01 of the spring peak value. The ISN~He haze spreads over large regions of the ram and anti-ram hemispheres and the completely dark portion of the sky (i.e., with the signal level below $10^{-4}$ of the peak) shows well only after the spring peak (orbit 72 and following). For the extreme case of kappa distribution, with $\kappa = 8/5$ and a large reference speed of $u_{\kappa} = 20.4$~\kms, even larger portion of the sky is occupied by the ISN~He haze (Figure~\ref{figMapE00kappa5}). As expected, in all simulated cases the fall peak is well-pronounced, located in the same location in the anti-ram hemisphere, and its height relative to the spring peak is approximately $2\%$. 

\begin{figure}
	\resizebox{\hsize}{!}{\includegraphics{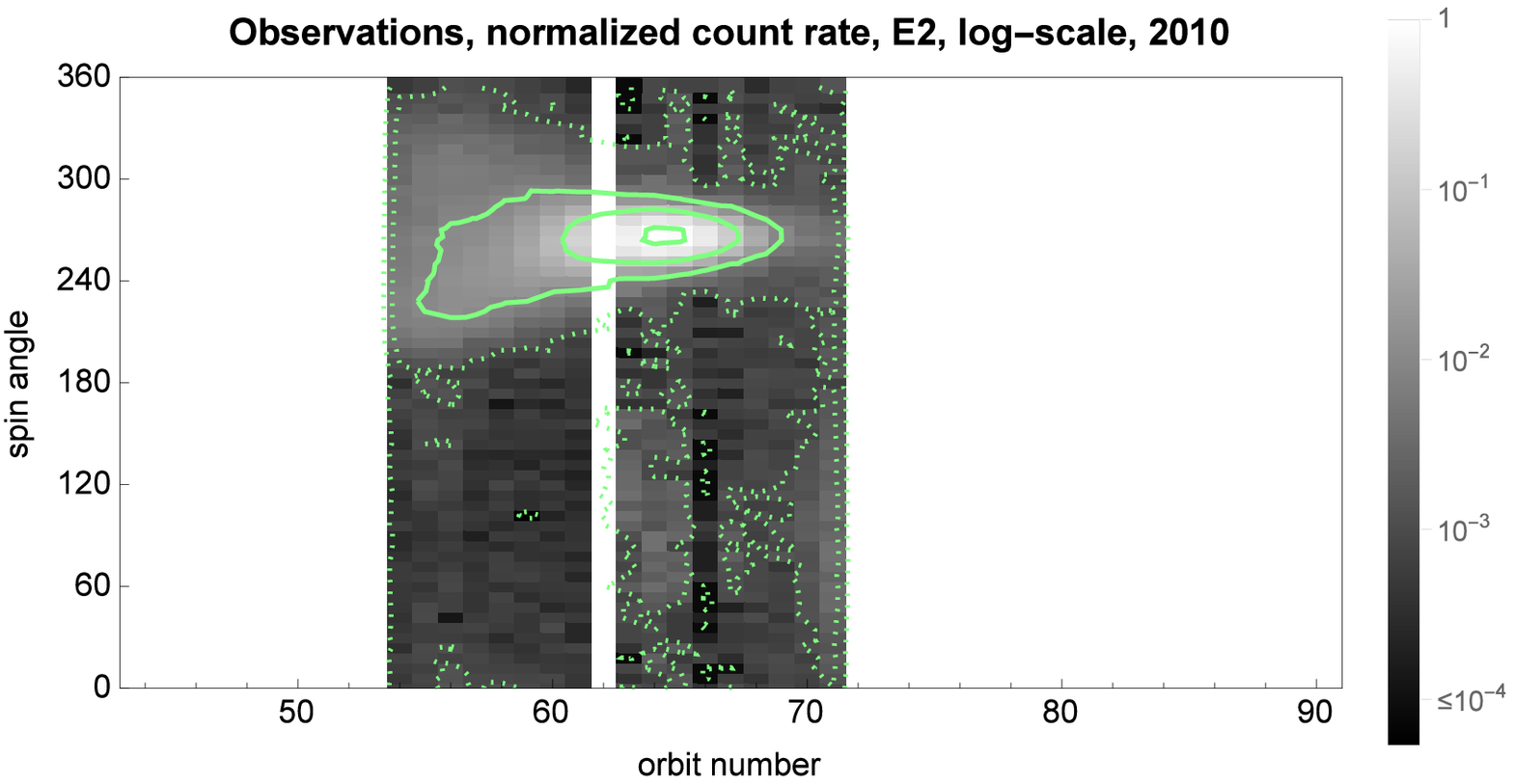}}
	\caption{Observations from energy step~2 of \emph{IBEX}-Lo taken in orbits 54 through 72. The grayscale shows the logarithm of the count rate normalized to the peak count rate registered during the 2010 ISN gas observations campaign (orbit 64). The isocontour lines mark the intensities of 0.8, $10^{-1}$, $10^{-2}$ (solid lines), $10^{-3}$ and $10^{-4}$ of the peak intensity (broken line). Data from orbit 62 are missing (white block).}
	\label{figMapE2obsWhite}
	\end{figure}

Inspection of Figure~\ref{figMapE2obsWhite}, which shows \emph{IBEX}-Lo measurements in energy step~2, taken for orbits 54 through 72, i.e., during the 2009/2010 ISN observation season, reveals a different picture. The data shown in this figure correspond to the data used by \citet{bzowski_etal:12a} to analyze the ISN~He population and by \citet{kubiak_etal:14a} to analyze the WB. They were meticulously cleaned from undesirable background and transients using a restrictive algorithm \citep{mobius_etal:12a, leonard_etal:15a} and represent the cleanest data set currently available for the studies of ISN~He gas. While the ISN~He peak and the WB region in the ram hemisphere are well visible, the entire anti-ram hemisphere is basically dark, with some transient islands of enhanced emission due to magnetospheric foreground \citep{galli_etal:14a}. The fall peak is not visible, since it is expected to occur a few orbits before the beginning of the ISN~He data. 

However, with the WB present and no energy threshold, one does expect some signal in the region that is apparently dark. The likely explanation for not observing this signal is an energy sensitivity threshold. \emph{IBEX}-Lo is sensitive to H atoms with energies above $\sim10$~eV. The sensitivity to He atoms is less well known, but the cutoff effect certainly exists, as illustrated by \citet{kubiak_etal:14a}, who showed in their Figures~8~and~9 that the wings of the WB beams in spin-angle change significantly with the assumed energy threshold. To verify this hypothesis, we simulated sky maps of the ISN~He flux seen by \emph{IBEX}-Lo assuming that the threshold values equal to 10, 19, and 38~eV, which approximately correspond to the lower boundary of \emph{IBEX}-Lo energy steps E1, E2, E3, respectively. The simulations do not include the process of the sputtering of the H atoms registered by \emph{IBEX}-Lo, but it is clear that a sputtered H atom cannot have an energy larger than the impacting He atom. Once sputtered, however, it can take any energy from a little below the impact energy down to zero, so it can be found in any \emph{IBEX}-Lo energy steps below the impactor's energy \citet[see more in][this volume]{galli_etal:15a}. 

\begin{figure}
	\resizebox{\hsize}{!}{\includegraphics{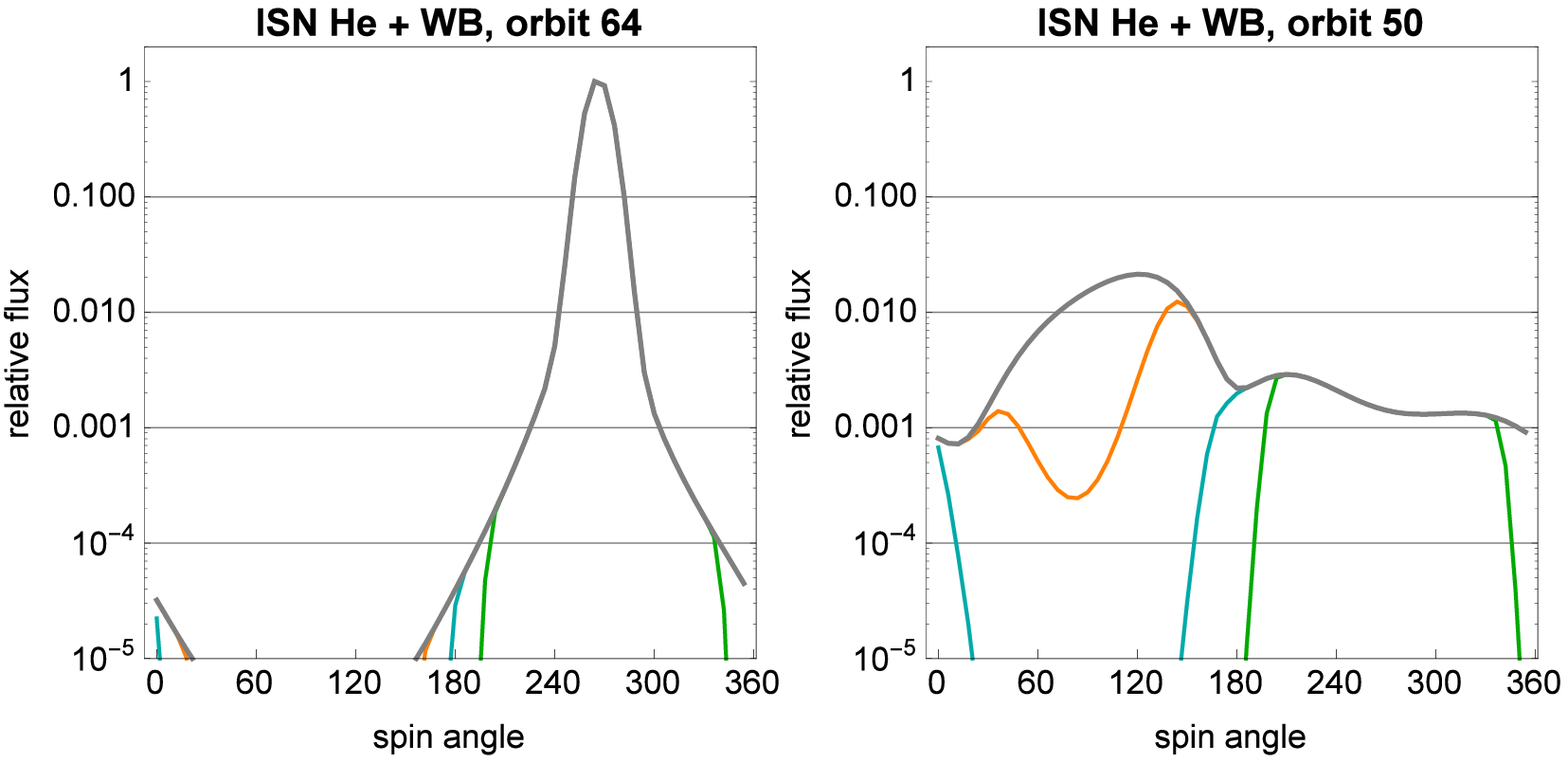}}
	\caption{He flux scaled to the maximum flux for orbits 64 (the spring peak orbit, left panel) and 50 (the fall peak orbit, right panel) for energy cutoff 0 (gray), 10~eV (orange), 19~eV (blue), and 38~eV (green), simulated assuming the ISN~He and Warm Breeze Maxwellian population in front of the heliopause.}
	\label{figNISWBLines}
	\end{figure}

\begin{figure}
	\resizebox{\hsize}{!}{\includegraphics{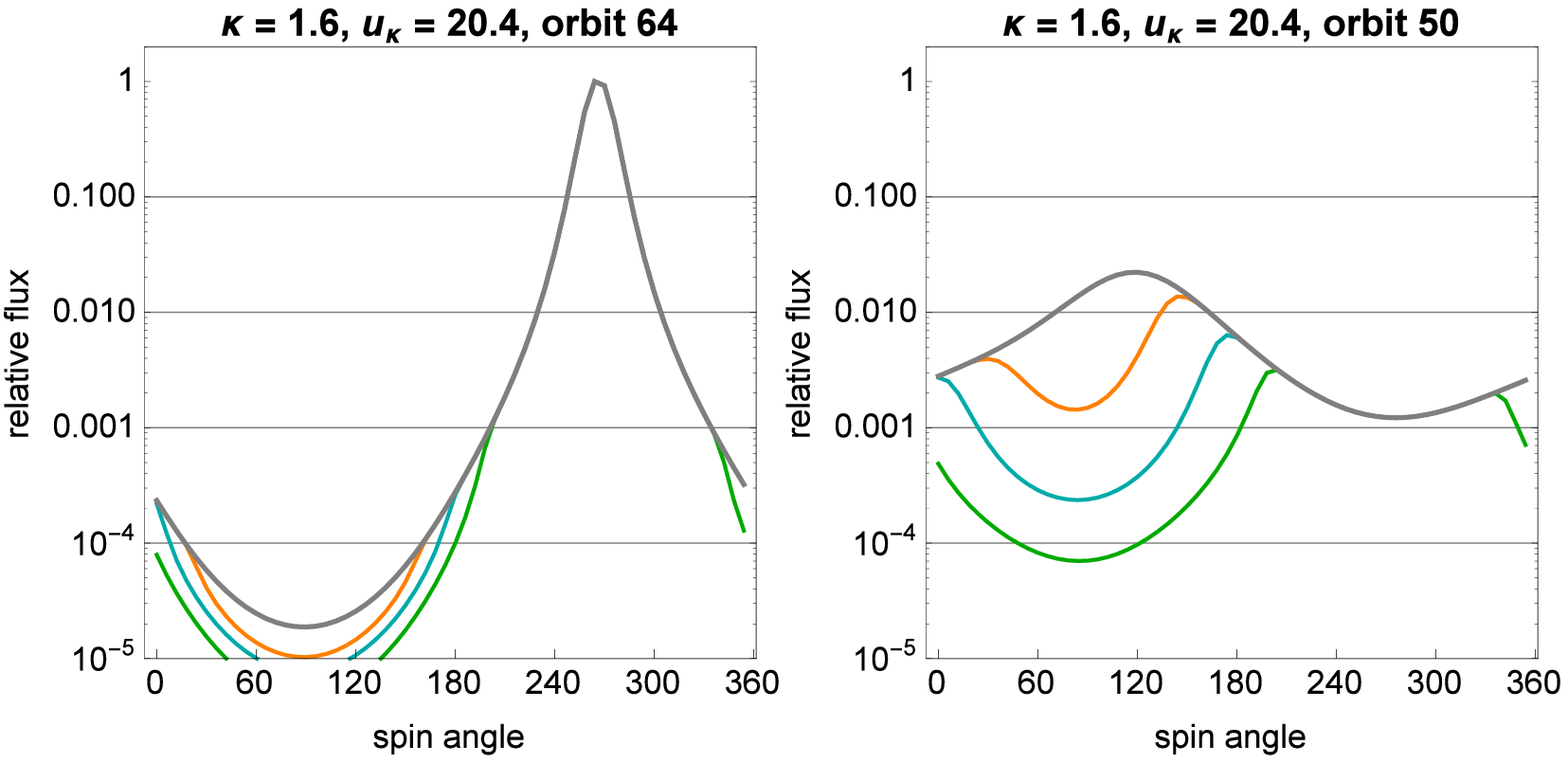}}
	\caption{He flux as in Figure~\ref{figNISWBLines}, calculated assuming a kappa population in front of the heliopause with $\kappa = 8/5$ and reference speed $u_{\kappa} = 20.4$~\kms. Note the profound difference between the signal in the anti-ram hemisphere, calculated for four different energy thresholds for the case of kappa function, presented in the right panel of this figure, with the signal simulated for identical energy thresholds for the case of the Maxwellian ISN~He + Warm Breeze, shown in the right panel of Figure~\ref{figNISWBLines}.}
	\label{figKappaLines}
	\end{figure}

\begin{figure}
	\resizebox{\hsize}{!}{\includegraphics{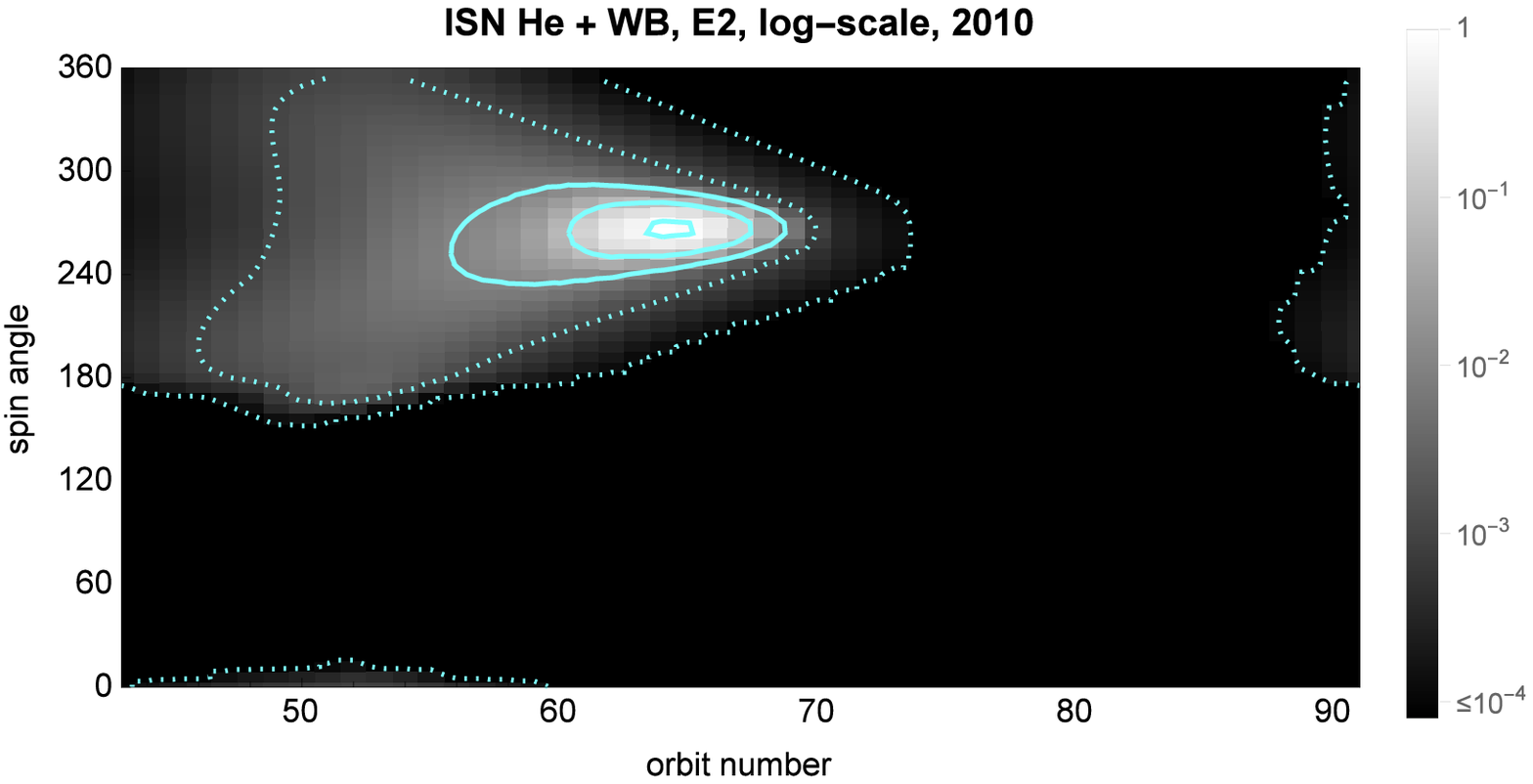}}
	\caption{Simulated sky map of helium flux (the ISN~He + Warm Breeze case), equivalent to Figure~\ref{figMapE00niswb}, calculated with the energy threshold of 19~eV.}
	\label{figMapE20niswb}
	\end{figure}

\begin{figure}
	\resizebox{\hsize}{!}{\includegraphics{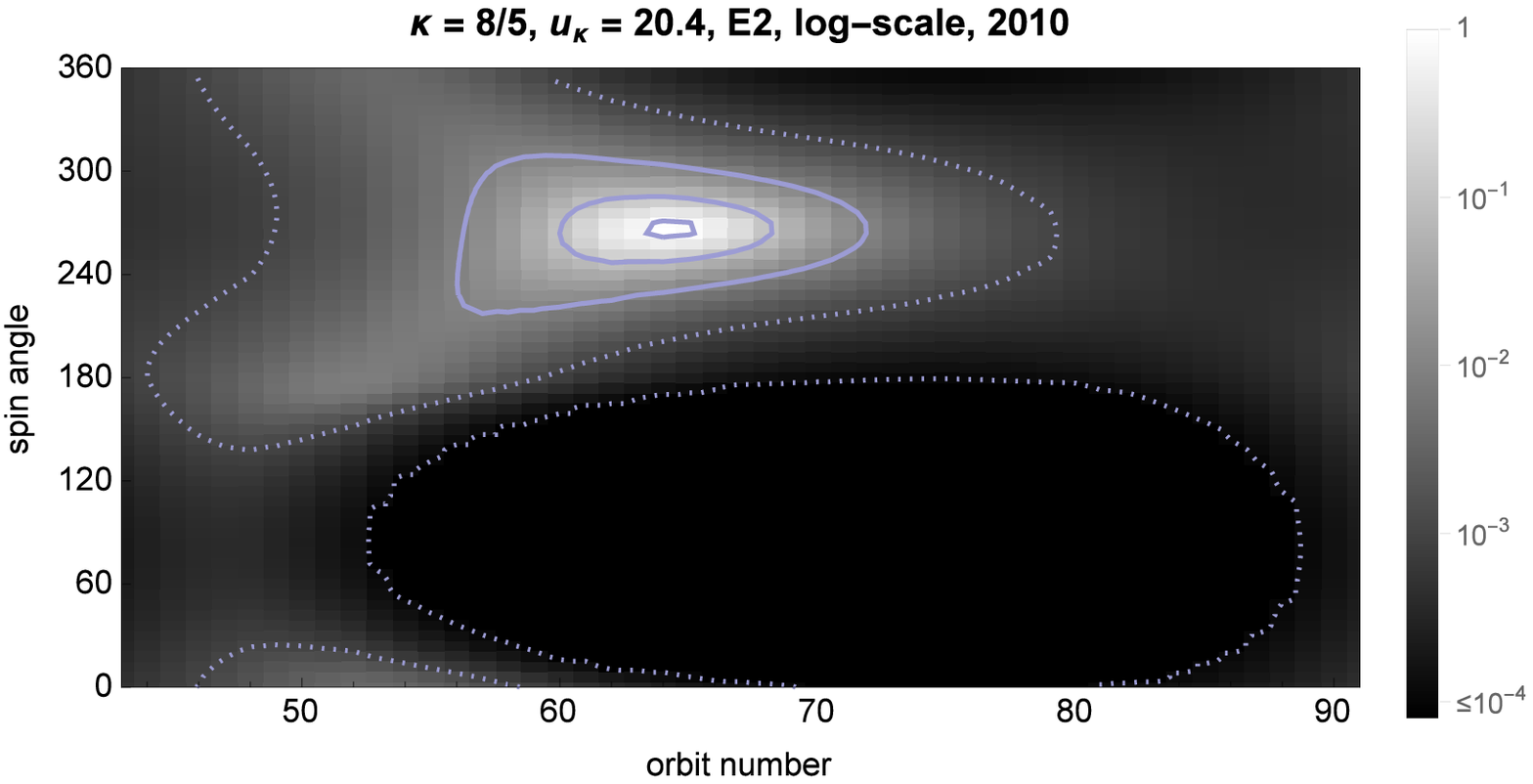}}
	\caption{Simulated sky map of ISN~He flux ($\kappa = 8/5$, $u_{\kappa} = 20.4$~\kms), equivalent to Figure~\ref{figMapE00kappa5}, calculated with the energy threshold of 19~eV.}
	\label{figMapE20kappa5}
	\end{figure}

The simulations show that the energy threshold has no effect on the flux registered in orbits where the spring peak is observed (see the left hand panels of Figures~\ref{figNISWBLines} and \ref{figKappaLines} for the ISN~He + WB and kappa function cases, respectively). This is because the relative energy of the atoms in the spacecraft frame is high due to the observations in the ram collision direction of detection. But for the orbits where the WB dominates and where the ISN~He haze is expected, the situation is different: the signal in the anti-ram hemisphere drops precipitously except for the case with the low threshold of 10~eV (see the right panels of Figures~\ref{figNISWBLines} and \ref{figKappaLines} and also Figures~8 and 9 in \citet{kubiak_etal:14a}). The fall peak is clearly visible only for the lowest-energy threshold, albeit shifted in spin-angle, and for higher-energy thresholds it is no longer visible. Generally, starting from a threshold of $\sim19$~eV the entire anti-ram hemisphere is expected to be almost totally dark, but the threshold has a relatively low impact on the signal in the ram hemisphere. Example sky maps for an energy threshold of 19~eV are shown in Figures~\ref{figMapE20niswb} and \ref{figMapE20kappa5} for the ISN~He + WB and kappa functions, respectively. Note that for the broad kappa function, the angular size of the spring peak is much larger than that for the ISN + WB Maxwellian case, which implies that the spring peak should be visible in the data for a greater portion of the year and reach to higher north and south latitudes.

\subsection{How to determine the energy threshold from observations: in search for the fall peak}
Determining the energy sensitivity threshold for neutral He observations by \emph{IBEX}-Lo is important for understanding the haze, both observed by \citet{fuselier_etal:14a} and \citet{galli_etal:14a}, and predicted by the simulations presented in this paper. As qualitatively shown in the previous section, the ratio of the fall peak to the spring peak is a function of the energy sensitivity threshold. Therefore we propose to examine the data in search for the fall peak. Since for the zero energy threshold the fall peak should exist at an intensity level of about $2\%$ of the spring peak intensity, which is well above the \emph{IBEX}-Lo background level, the lack of a fall peak in the data will be evidence that there is a finite energy cutoff in the instrument sensitivity to neutral He. To facilitate finding the likely value of this cutoff and to find at which energy threshold the fall peak disappears in the background, we calculated the behavior of the fall peak for the distribution functions and gas parameters discussed earlier at a few adjacent orbits, with an energy threshold increasing from 0 to $\sim38$~eV. We performed the calculations for the conditions of the 2009/2010 and 2013/2014 observation seasons. 
	\begin{figure}
	\resizebox{\hsize}{!}{\includegraphics{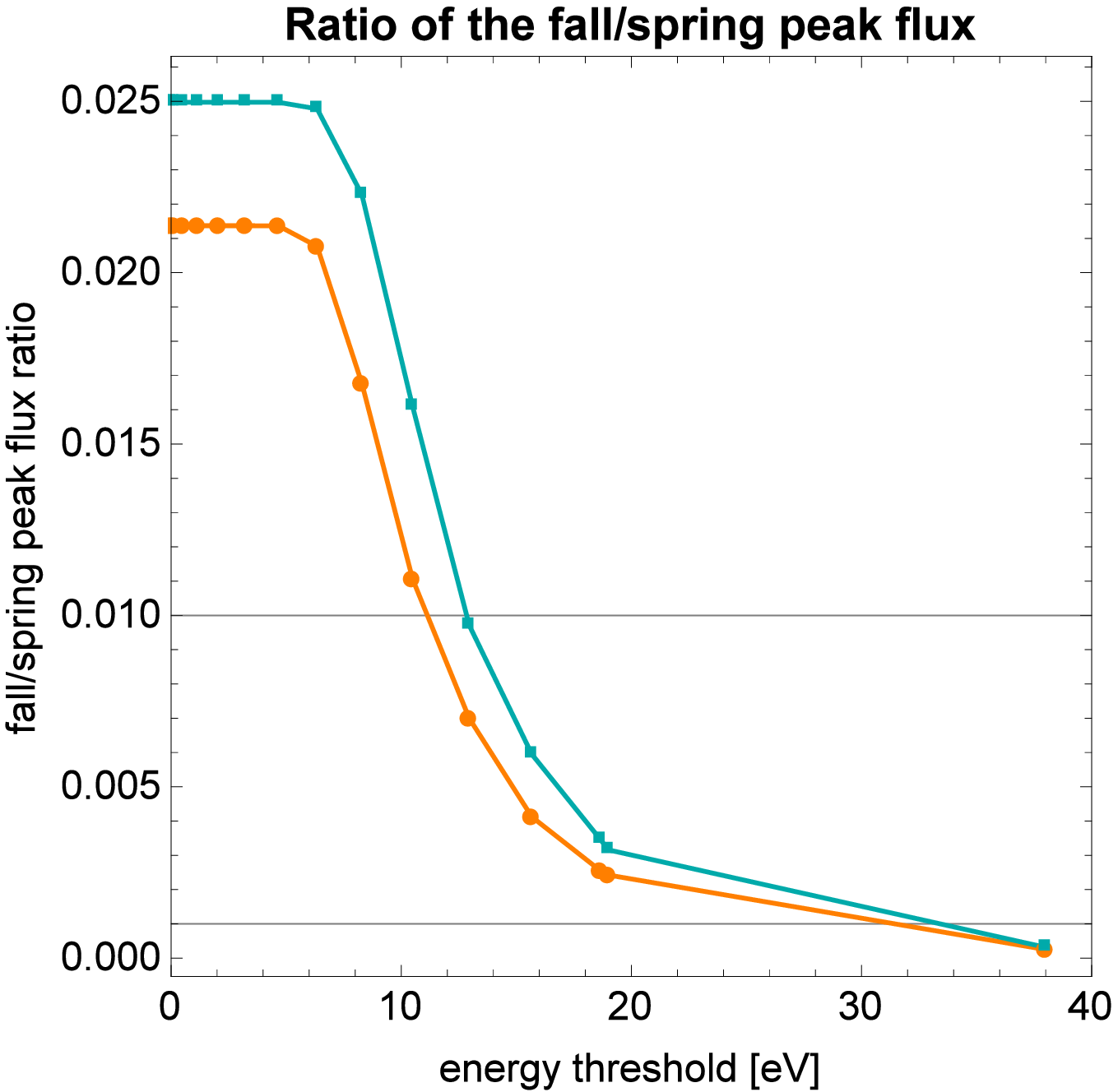}}
	\caption{Ratios of the fluxes in the highest pixels in the fall and spring peaks for the 2009/2010 (orange) and 2013/2014 (blue) ISN~He observation seasons, shown as a function of the energy cutoff, for the single-Maxwellian distribution function of ISN~He. The horizontal bars correspond to the $10^{-3}$ and $10^{-2}$ isocontours shown in Figure~\ref{figMapE00nis}.}
	\label{figPeakVsEnergy}
	\end{figure}

The results are shown in Figures~\ref{figNISWBLines}, \ref{figKappaLines}, and \ref{figPeakVsEnergy}. The fall peak is expected to be unaffected by the energy threshold below $\sim7$~eV. For larger values, the fall peak is rapidly reduced and drops to a half of its value for the threshold at $\sim12$~eV, and for $\sim19$~eV it is reduced to $\sim0.002$, i.e., the peak is effectively gone. For the kappa function and ISN~He + WB cases, the drop with energy is less steep but the $\sim19$~eV limit for suppressing the fall peak is supported. The spin-angle of the peak varies with the threshold energy, moving northward for increasing threshold values. Not surprisingly, the profiles of the peak depend on the assumed distribution function, but the relative fall/spring peak height is affected only weakly. We checked that the increased ionization, characteristic for the current maximum of solar activity \citep{sokol_bzowski:14a}, does not modify the conclusions.

The experimental search of the threshold energy for the \emph{IBEX}-Lo detector is presented in the associated paper by \citet[][this volume]{galli_etal:15a}. Their analysis of \emph{IBEX}-Lo observations suggests that there exists an energy cutoff for ISN~He detection and that it lies between 25 and 30~eV for hydrogen sputtered by helium. For a thorough discussion of the search of fall peak in the \emph{IBEX}-Lo data we refer the reader to the paper by \citet[][this volume]{galli_etal:15a}.

\subsection{Where to look to discriminate between the kappa function and two Maxwellians}
We find that the simulated sky maps for the two cases of distribution functions of the ISN~He gas in front of the heliopause (Maxwellian primary ISN He + Maxwellian WB, and single-population kappa distribution function) show very distinct differences in the region just to the right of the spring peak (see Figures~\ref{figMapE00niswb} and \ref{figMapE00kappa5}) and these regions may be crucial for discriminating between these two distributions of ISN~He at the source region. The span of the region where the ISN haze flux exceeds the $10^{-3}$ isocontour is much wider in the case of the single kappa population than for the two-Maxwellian case. This difference holds regardless of the magnitude of the energy threshold, as can be verified by comparing simulated sky maps plotted for these two cases for the likely energy threshold of $\sim19$~eV (Figures~\ref{figMapE20niswb} and \ref{figMapE20kappa5}).

\begin{figure}
	\centering
	\begin{tabular}{c}
	\includegraphics[scale=0.6]{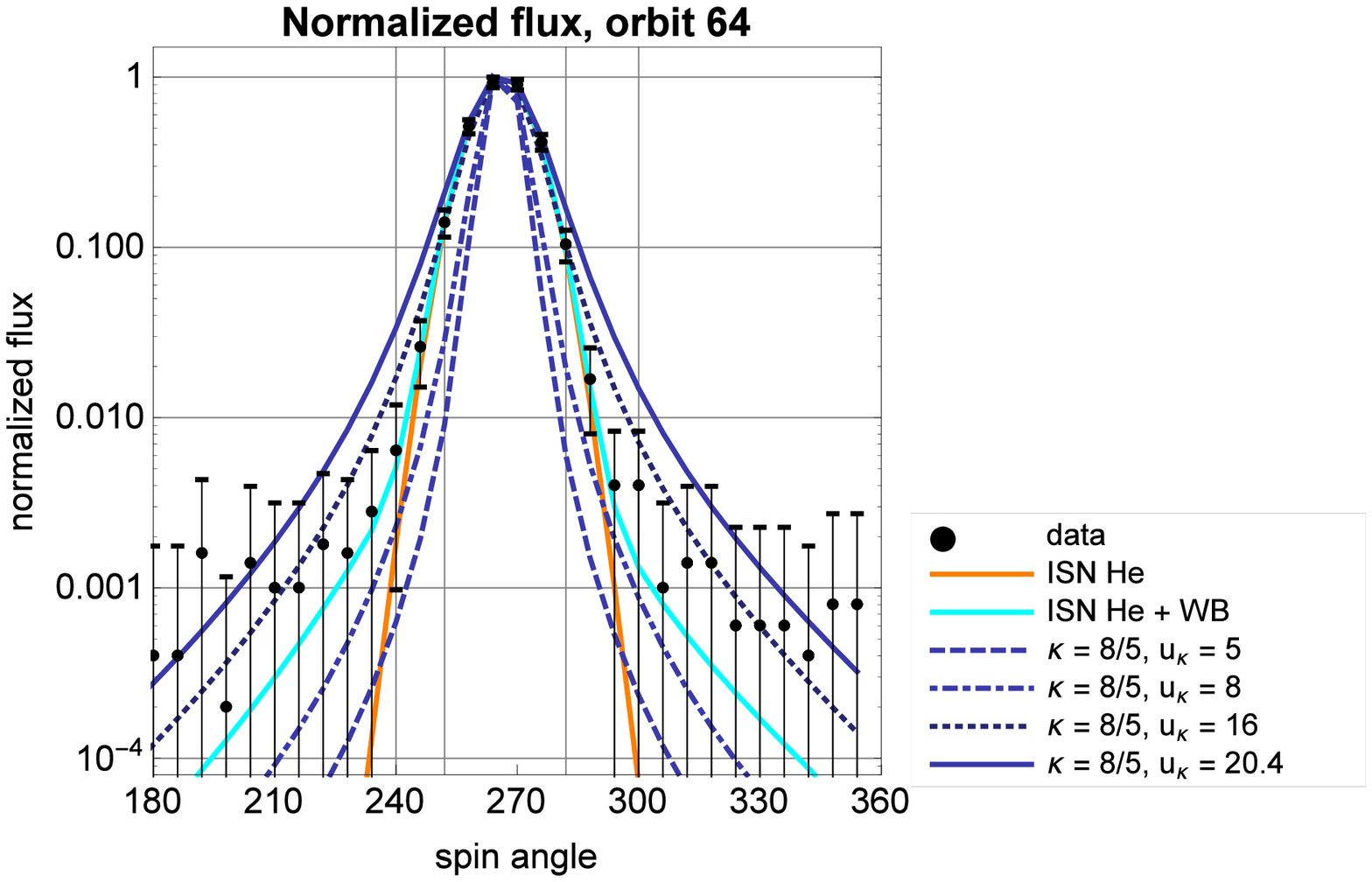}\\
	\includegraphics[scale=0.6]{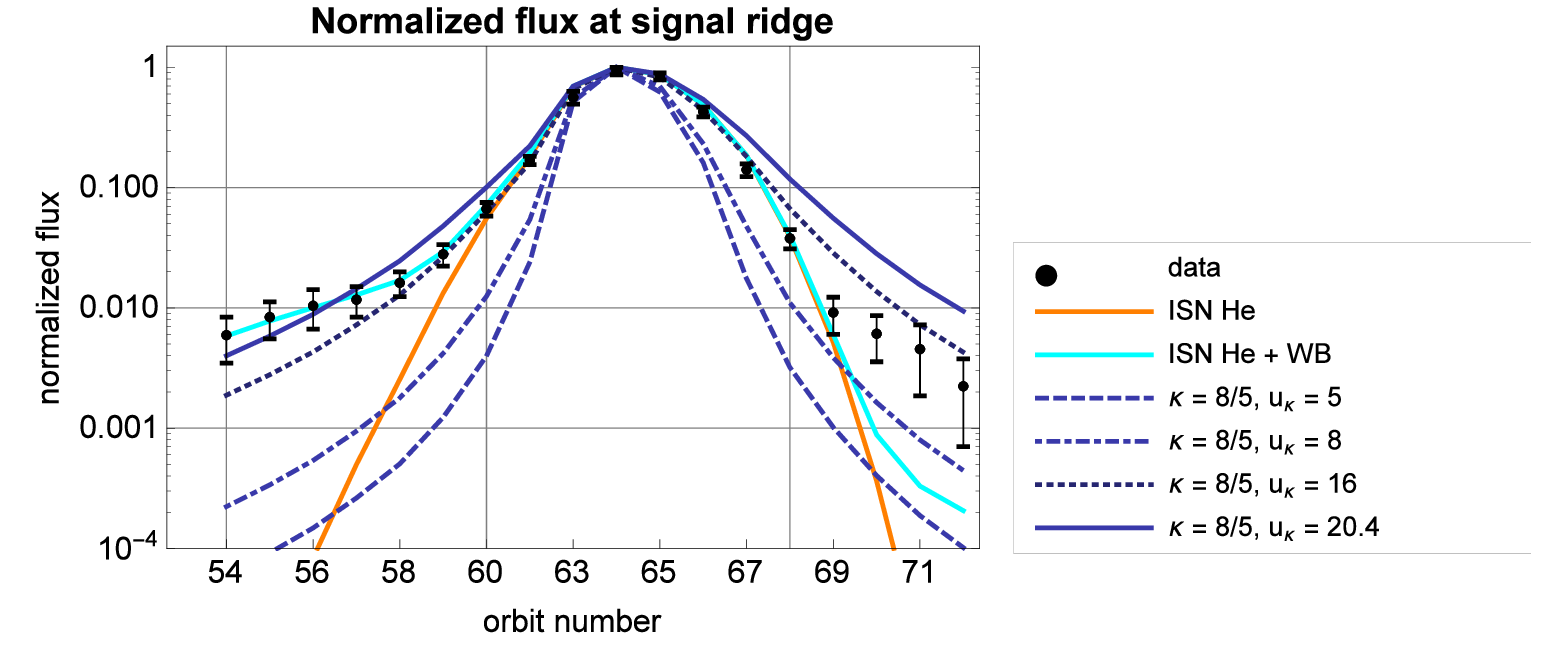}\\
	\end{tabular}
	\caption{Two cuts through the sky maps discussed in the paper. The upper panel shows the cut along the spin angle for orbit 64 (spring peak, $\lambda_{E}\sim 136\degr$, the vertical direction in the sky map plots) and the lower panel shows the cut along a fixed spin angle value of $264\degr$, equal to the spin angle of the maximum flux in orbit 64. The colors are explained in the legend, and $u_{\kappa}$ is given in \kms. The innermost vertical grid lines in the upper panel mark the data points taken by \citet{bzowski_etal:12a} for ISN~He analysis, and the outermost grid lines the data points used by \citet{kubiak_etal:14a} for the Warm Breeze analysis. The extreme left and right vertical grid lines for orbits 54 and 68, respectively, in the lower panel mark the first and the last orbit used by \citet{kubiak_etal:14a} for the Warm Breeze analysis, while the grid lines 61 and 68 span the subset of orbits used by \citet{bzowski_etal:12a} for ISN~He analysis.}
	\label{figSimVsData}
	\end{figure}

We plot two cuts through the simulated and observed maps to better illustrate the differences between the discussed cases. One cut goes along the spin-angle (effectively, a cut in ecliptic latitude) on the spring peak orbit 64 (Figure~\ref{figSimVsData}, upper panel), the other one has a fixed spin-angle corresponding to the spring peak and runs through all of the orbits (Figure~\ref{figSimVsData}, lower panel), which is equivalent to scanning the flux in ecliptic longitude along the signal ridge. We use simulations for the zero energy threshold but the flux measured in these regions of the \emph{IBEX} sky is not sensitive to the energy threshold (this insensitivity is demonstrated in Figures~\ref{figNISWBLines} and \ref{figKappaLines}). 
	
It is clearly seen from the comparison of the data with models that the two-Maxwellian case of ISN~He + WB fits the data almost perfectly, but this is understandable since the parameters of the latter case were obtained from fitting performed by \citet{bzowski_etal:12a} and \citet{kubiak_etal:14a}. In contrast, in the extreme case of a single-population kappa functions with $\kappa = 8/5$, neither $u_{\kappa} = 20.4$~\kms nor $u_{\kappa} = 5$~\kms fit to the data even qualitatively. Also the kappa cases with intermediate reference speeds $u_{\kappa}$ of 8 and 16~\kms do not fit. Admittedly, the $\kappa$ and $u_{\kappa}$ values were taken arbitrarily, but it seems that the differences are fundamental and most likely impossible to remove by adjusting these parameters. The narrow version of the kappa function (i.e., with a low $u_{\kappa}$ value) is much too narrow, and the kappa function at the other extreme ($u_{\kappa} = 20.4$~\kms) is too wide in ecliptic latitude. The orbital peaks of the flux given by the wide kappa function fit the observed peaks with only small deviations, but with a very notable exception of orbits immediately after the spring peak, i.e., orbit 66 and subsequent ones. Perhaps one could think of increasing the kappa value while reducing the reference speed $u_{\kappa}$, but there seems to be little room for adjustments that would not break the reasonable agreement between the data and the simulation seen in the left branch in the lower panel of Figure~\ref{figSimVsData}.

The data from the region observed in orbits after the spring peak seem to be at odds with a single wide kappa function; with $\kappa = 8/5$, the predicted signal is well above the data. The simulated signal for the ISN + WB case is below the observed signal, but one needs to keep in mind that this region has an appreciable contribution from ISN~H \citep{saul_etal:12a, saul_etal:13a, schwadron_etal:13a}. It seems that the crucial regions for looking for evidence of non-equilibrium distribution of neutral He in front of the heliopause will be the orbits after the spring peak each year, and within these orbits, the pixels are at a distance of $\sim24\degr$. Pixels farther away have poor statistics and seem to be a mixture of the ISN~He haze and background counts in proportions unknown \emph{a priori}. Reducing the background and increasing the statistics is important to better characterize departures of the neutral He gas from equilibrium in front of the heliosphere. 

Looking for evidence for non-equilibrium distribution functions of ISN~He gas in front of the heliopause requires finding the energy sensitivity threshold of \emph{IBEX}-Lo. When this threshold is established from the analysis of observations for fall peak orbits, it will be possible to test if a kappa distribution function, with $\kappa$ and $u_{\kappa}$ parameters found from fitting the data, better describes the observations.

\section{Summary and conclusions}
In the analysis of the ISN~He flux from the first two years of \emph{IBEX} data collection \citet{bzowski_etal:12a} reported on the elevated wings of the observed flux, which cannot be explained by a single Maxwell--Boltzmann distribution of the gas in the source  region. \citet{gruntman:13a} tried to explain this additional signal as being due to the elastic collision of the He atoms with the solar wind protons. \citet{kubiak_etal:14a} found that a part of this flux can be explained by the additional ISN He population, dubbed the Warm Breeze. In the meantime, \citet{galli_etal:14a} and \citet{fuselier_etal:14a} reported on a ubiquitous ``background'' visible by \emph{IBEX} in the lowest energies. These findings motivated us to check if the additional signal can be described by the weak ISN He flux that should be distributed on the whole sky. We call it the ISN He haze due to its weakness and ubiquitousness. The haze should be present regardless of the distribution function of the gas and the nature of the WB. In addition to the ISN~He haze, and in addition to the spring peak of ISN~He gas, observed by \emph{IBEX} during the first quarter of each year, there must be the fall peak, potentially observable when the spacecraft is located at the opposing side of the downwind axis. This fall peak is reduced in intensity by a factor of 50 relative to the spring peak due to the Compton--Getting effect (Figure~\ref{figPeakVsEnergy}). 

We show that without an energy threshold, ISN~He haze should be observed in both ram and anti-ram hemispheres. However, the intensity of the haze, as well as the fall peak height and location, are sensitive functions of the energy threshold of the instrument. The fall peak is expected in orbits from 49 to 51 ($\lambda_{E}$ from $\sim20\degr$ to $\sim40\degr$) and equivalent in other observation seasons at a level exceeding the \emph{IBEX}-Lo background level. A careful analysis of data from these orbits should either reveal the peak or show that it is not there, which should be interpreted as independent evidence for the existence of an energy threshold in the \emph{IBEX}-Lo sensitivity. We show that a non-detection of the fall peak would suggest that the energy threshold is at a level of at least $\sim19$~eV (Figure~\ref{figPeakVsEnergy}). Similar conclusions are presented in the accompanying paper by \citet[][this volume]{galli_etal:15a}. 

We present full-sky maps of the expected ISN~He haze in both ram and anti-ram hemispheres for various energy thresholds and for various models of the distribution function of ISN~He in front of the heliopause: one-population Maxwell--Boltzmann (Figure~\ref{figMapE00nis}), corresponding to the primary ISN~He population, convolution of two populations (it is primary ISN He and WB) given by  Maxwell--Boltzmann distribution function, or kappa populations with a low $\kappa$ index of $8/5$ and a number of reference speed $u_{\kappa}$ values, ranging from a low $u_{\kappa} = 5$~\kms (Figure~\ref{figMapE00kappa1}) to an extremely and unrealistic high $u_{\kappa}\sim20.4$~\kms (Figure~\ref{figMapE00kappa5}). 

We show that these alternative hypotheses will create very different ISN~He hazes and that ISN~He maps from \emph{IBEX}-Lo, once the energy sensitivity threshold is established, can be used to differentiate between them. Most likely, the low value of $\kappa = 8/5$ is not supported by the data, but it is feasible that a kappa population with relatively low kappa value is a better model for the WB than the Maxwell--Boltzmann distribution of gas. 

\acknowledgments
The authors from SRC~PAS were supported by Polish National Science Centre grant 2012-06-M-ST9-00455. A.G. and P.W. acknowledge the financial support by the Swiss National Science Foundation. E.M., H.K., S.F., and D.M. were supported by the IBEX mission as a part of the NASA Explorer Program.

\bibliographystyle{apj}
\bibliography{iplbib}{}

\end{document}